\documentclass[a4paper,aps,prd,twocolumn,eqsecnum,amsmath,amssymb,nofootinbib,superscriptaddress]{revtex4-2}
\usepackage{dcolumn}
\usepackage{bm}
\usepackage{graphics}
\usepackage{graphicx}
\usepackage{epsfig}
\usepackage{booktabs,multirow}
\usepackage{hhline}
\usepackage{slashed}
\usepackage{rccol}
\usepackage{mathrsfs}
\usepackage{amsmath, amssymb}
\usepackage[dvipsnames]{xcolor,colortbl}
\usepackage{subfigure}
\graphicspath{{figures/}}

\usepackage{xcolor}\colorlet{linkequation}{blue}
\usepackage[colorlinks=true, % false: boxed links; true: colored links
linkcolor=Red,   % color of internal links
filecolor=Red,   % color of file links
citecolor=Green, % color of links to bibliography
urlcolor=Blue,
hyperfootnotes=true]{hyperref}

\newcommand*{\SavedEqref}{}
\let\SavedEqref\eqref
\renewcommand*{\eqref}[1]{%
	\begingroup
	\hypersetup{
		linkcolor=linkequation,
		linkbordercolor=linkequation,
	}%
	\SavedEqref{#1}%
	\endgroup
}

\begin{document}
\title{Probing light mediators with recent PandaX-4T low energy electron recoil data}
%%Begin Author List
\newcommand{\ktu}{Department of Physics,
Karadeniz Technical University, Trabzon, TR61080, Türkiye}
\newcommand{\ktufb}{Graduate School of Natural and Applied Science, Karadeniz Technical University, Trabzon, TR61080, Türkiye}
\author{M.~Demirci}
\email{mehmetdemirci@ktu.edu.tr (corresponding author)}
\affiliation{\ktu}

\author{M.~F.~Mustamin}
\email{mfmustamin@ktu.edu.tr}
\affiliation{\ktu}
%%End Author List

\date{\today}

\begin{abstract}
Neutrinos elastically scattered off atomic electrons offer a unique opportunity to probe the Standard Model (SM) and beyond SM physics. In this work, we examine the new physics effects of light mediators through elastic neutrino-electron scattering using solar neutrinos at the low energy range of PandaX-4T. These mediators, with a mass less than $1$ GeV, are common properties of extensions to the SM.
Accordingly, we consider universal light mediator models involving scalar, vector, and tensor interactions allowed by Lorentz invariance, and the anomaly-free $U(1)'$ extensions of the SM with a new vector mediator such as the $L_e-L_\mu$, $L_e-L_\tau$ and $L_\mu-L_\tau$ gauge models in which the charges are exclusively leptonic.  
The new physics effects are analyzed by embedding each model into the SM process using solar neutrino flux. We obtain novel constraints on the coupling-mass plane of these models from the latest Run0 and Run1 datasets of the PandaX-4T experiment. We also compare our results with other limits derived from various terrestrial and astrophysical experiments. Our analysis results reveal more stringent limits in some regions of parameter space.
\end{abstract}

%\medskip

\maketitle
\section{Introduction}
Elastic neutrino-electron scattering is a precisely known pure leptonic process in which neutrinos scatter off an electron \cite{tHooft:1971ucy}. This reaction proceeds through the charged current (CC) and neutral current (NC) interactions. This makes it a convenient channel for testing the electroweak sector of the Standard Model (SM) with a clean experimental signature. In connection with the solar neutrino, the scattered electrons are primarily in the forward direction along the Earth-Sun vector \cite{Bahcall:1995mm} and are sensitive to different neutrino flavors.

The solar neutrino is one of the most intensive natural neutrino sources on Earth. It is produced as an electron neutrino through fusion processes in the core of the Sun and could change flavor during propagation.. The subject has been widely worked on since its first observation \cite{Davis:1968cp}. Solar neutrino measurements are closely related to the discovery of flavor conversion and the effect of matter on neutrino dispersion, which are the results of some channels, such as CC \cite{GALLEX:1992gcp, Cleveland:1998nv, SAGE:1999uje}, NC \cite{SNO:2002tuh, SNO:2003bmh, SNO:2008gqy}, and elastic scattering \cite{Kamiokande-II:1989hkh, Super-Kamiokande:2001ljr, Borexino:2007kvk, Borexino:2008fkj}, which lead to the formulation of the Standard Solar Model \cite{Bahcall:2000nu, Bahcall:2004mq, Bahcall:2004pz, Serenelli:2016dgz, Vinyoles:2016djt, Vitagliano:2019yzm}. 

On the other hand, solar neutrinos can induce both elastic neutrino-electron and coherent neutrino-nucleus scatterings in dark matter (DM) direct detection (DD) experiments, resulting in detectable event rates at current facilities \cite{Cerdeno:2016sfi}. 
They are part of the irreducible background signals in the experiments. Direct detection of DM related to neutrinos from celestial sources was proposed in the mid-eighties \cite{Drukier:1984vhf, Goodman:1984dc, Drukier:1986tm}. Nowadays, these searches are ongoing in numerous experiments such as CDEX \cite{CDEX:2022mlp}, XENON \cite{XENON:2020kmp, XENON:2022ltv}, LUX-ZEPLIN (LZ) \cite{LUX:2015abn, LZ:2018qzl,LZ:2023ja}, DarkSide \cite{DarkSide:2018kuk}, and PandaX \cite{PandaX:2014mem, PandaX-II:2017hlx}. 
Furthermore, the next generation of DD experiments has intensified programs on DM and neutrino physics with multi-ton liquid-xenon (LXe) and argon (LAr) detectors such as DARWIN \cite{DARWIN:2020bnc} and DarkSide-20k (DS-20k) \cite{DARKSIDE20K:2021}. Recently, the solar $^8$B neutrino flux has been measured through coherent elastic neutrino-nucleus scattering (CE$\nu$NS) in the PandaX-4T \cite{PandaXT:20248b} and XENONnT \cite{XENONnT:20248b} dark matter experiments with a statical significance of 2.64$\sigma$ and 2.73$\sigma$, respectively. This demonstrates the potential of xenon direct detection experiments to supply precise measurements of solar neutrino fluxes. For many years, such experiments aimed to search for DM-nucleon scattering; however, as detectors were improved and low-background events were better understood, electron recoils became an interesting framework to probe light DM \cite{XENON:2022ltv,LZ:2023ja, PandaX:2022ood}. The PandaX-4T experiment \cite{PandaX:2018wtu,PandaX:2024zbo,PandaX:2024cic,PandaX:2022ood,PandaX:2024zbo,PandaX:2024med} has such a detector. 
The LXe time projection chamber of the experiment was originally designed to observe light DM, but due to its large volume, low threshold, and low background ratio, it is also then used to search for new light particles and neutrino physics.

It is feasible to see the SM as an effective theory since it excellently describes behaviors of fundamental particles and that new physics effects are suppressed at low energies. In many extensions beyond the SM (BSM), low-mass particles appearing from hidden sectors are widely used to incorporate this presumption. Examples include a light DM candidate \cite{Arkani-Hamed:2008hhe}, sterile neutrinos that explain the non-zero mass of neutrinos \cite{Dasgupta:2021ies}, or a mediator for a dark fermion proposal \cite{Brdar:2018}. Many experiments are currently being developed to study such models (see, e.g., the review \cite{Essig:2013lka}). 
In particular, DD experiments allow us to investigate contributions from such new theories with light mediators owing to neutrino-electron scattering events induced by solar neutrinos.
Because this pure leptonic scattering process is well predicted in the SM, a measured deviation can strongly indicate signatures of BSM physics.

In the present work, with the above motivations, we study the new physics effects of the light mediator models on neutrino-electron scattering using measurements of the solar neutrino flux in the light of recent data of DD experiments.
We focus on the universal light mediator \cite{Cerdeno:2016sfi, Abdallah:2015ter}
and the lepton flavor-dependent $U(1)'$ \cite{Mohapatra:1980qe,He:1991qd} models. The former ones are constructed with the universal scalar, vector, and tensor interactions allowed by Lorentz invariance.  
The latter ones are particularly obey the anomaly-free $U(1)_{L_e-L_{\mu}}$, $U(1)_{L_e-L_{\tau}}$ and $U(1)_{L_\mu-L_{\tau}}$ gauge models (so-called leptophilic $Z'$ models) \cite{He:1991qd} in which the gauge charges are exclusively leptonic. These include a vector mediator that is coupled to the SM leptons, where the couplings differ from the universal vector case according to their charges.
All of these are theoretically well motivated to provide a complementary explanation to a series of emerging discrepancies in precision studies of low-energy observables. 
Such light mediators have previously been studied in the framework of both neutrino-electron and coherent elastic neutrino-nucleus scattering events from nuclear reactors \cite{CONUS:2021dwh, Melas:2023olz, Lindner:2024eng, Chattaraj:2025}, stopped-pion source \cite{Farzan:2018gtr, Cadeddu:2020nbr, Demirci:2021zci, DeRomeri:2022twg,  AtzoriCorona:2022moj, Coloma:2022avw}, and also solar neutrino \cite{Coloma:2022umy, Coloma:2020gfv, Gninenko:2020xys} at DD experiments \cite{Boehm:2020ltd, Schwemberger:2022fjl, A:2022acy, Khan:2023b, Demirci:2023tui, DeRomeri:2024dbv} as well as other experiments \cite{Bauer:2018onh, Lindner:2020kko, Bilmis:2015lja}. In particular, extra vector mediators have been widely studied, as they can explain some problems in SM such as baryogenesis through leptogenesis \cite{Fukugita:1986hr}, grand unified theory \cite{Buchmuller:1991ce}, nature of DM \cite{Alves:2015pea}, and anomalies from experiments \cite{Allanach:2015gkd, Allanach:2023uxz}.

The PandaX-4T collaboration has recently reported electron recoil signals with unblinding of the Run0 and Run1 data \cite{PandaX:2024zbo, PandaX:2024cic}.   
In this work, we derive novel limits on the coupling-mass parameters of the universal light mediator and the lepton flavor-dependent $U(1)'$ models using both datasets for the first time. Additionally, we present constraints from the latest electronic recoil XENONnT data \cite{XENON:2022ltv} for the lepton flavor-dependent $U(1)'$ models. There are no individual XENONnT results in the literature for $L_e-L_\mu$ and $L_e-L_\tau$, but it is derived for $L_\mu-L_\tau$ by Ref. \cite{Melas:2023olz} and our results are in excellent agreement with their results.
We compare our results with the available previous limits derived from the DD experiments (XENONnT, CDEX-10 \cite{Demirci:2023tui}) and solar neutrino facility (BOREXINO) mentioned previously, as well as from stopped-pion source (COHERENT \cite{COHERENT:2017ipa, COHERENT:2020iec, COHERENT:2021xmm}), nuclear reactor (GEMMA \cite{Beda:2009kx}, TEXONO \cite{TEXONO:2009knm}, CONNIE \cite{CONNIE:2019xid}, CONUS \cite{CONUS:2020skt}, CONUS+ \cite{CONUSplus:2025}, Dresden-II \cite{Colaresi:2022obx}), and beam dump experiments (CHARM-II \cite{CHARM-II:1994dzw}, LSND \cite{LSND:2001akn}, NA64 \cite{NA64:2022yly}). We also include limits obtained from the collider experiments (Mainz \cite{A1:2011yso}, KLOE \cite{ALICE:2012aqc}, BaBar \cite{BaBar:2014zli}, PHENIX \cite{PHENIX:2014duq}) %LHCb \cite{LHCb:2019vmc}) 
and rare-meson decay (NA48/2 \cite{NA482:2015wmo}). We further complement our results with limits from the $(g-2)_\mu$ \cite{Muong-2:2023cdq} as well as astrophysical sources \cite{Blinov:2019gcj, Heurtier:2017, Huang:2018, Croon:2020lrf, Escudero:2019gzq,LiXu:2023}.

We organize the remainder of our paper as follows. In Sec.~\ref{sec:nue}, we introduce the theoretical formulation of the neutrino-electron scattering process both in the SM and in the presence of new light mediators. We also present the calculation details of the differential spectra. In Sec.~\ref{sec:stat}, we introduce the data analysis method used in the limit setting, highlighting all details. In Sec.~\ref{sec:resdis}, we present expected event spectra for each model. Then we provide new constraints on the allowed parameter space and compare them with the other available constraints. Finally, we summarize and conclude our results in Sec.~\ref{sec:summ}.

\section{General Framework}\label{sec:nue}
In this section, we present the necessary pieces for event-rate calculations of solar neutrinos scattering elastically off electrons in the PandaX-4T detector; specifically, we discuss interaction cross sections in the SM and beyond in Sec.~\ref{sec:SM} and \ref{sec:BSM}, respectively, and calculations of the expected event rate in Sec.~\ref{sec:ER}.

\subsection{Standard formalism of neutrino-electron scattering}\label{sec:SM}
The elastic neutrino-electron scattering is a purely leptonic process in SM. In this process, the neutrino scatters off an electron by the exchange of a charged boson (only $\nu_e$) or a neutral boson (all flavors). 
\begin{figure}[h]
	\centering
	\includegraphics[scale=0.61]{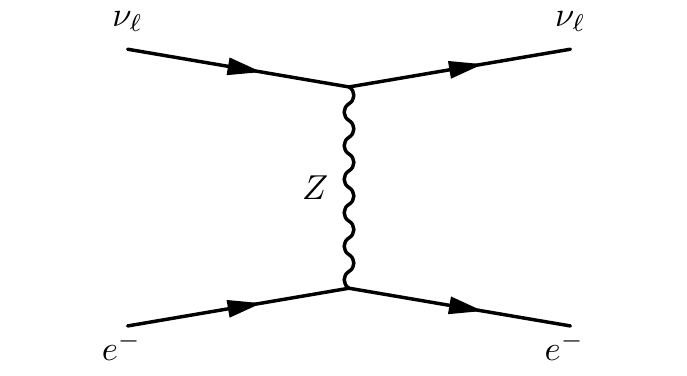}
	\includegraphics[scale=0.61]{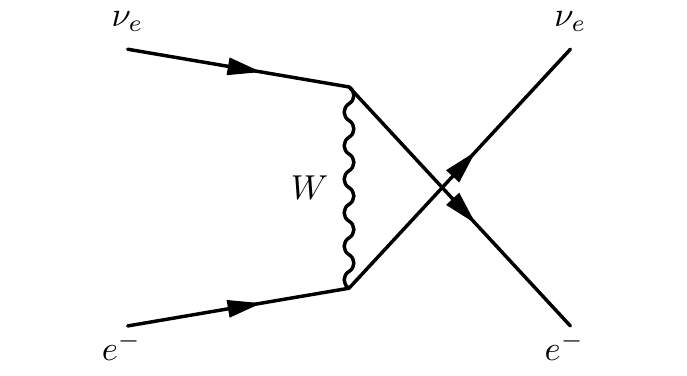}
	\\
	(a) \hspace{35mm} (b)
	\\
	\caption{Feynman diagrams for neutrino-electron scattering via (a) NC and (b) CC channels in the SM. The index $\ell$ runs for $e,\mu$, and $\tau$ flavor of neutrino.}
	\label{fig:nue_diag}
\end{figure}
The corresponding Feynman diagrams at tree level are shown in Fig.~\ref{fig:nue_diag}. Here, $Z$ and $W$ represent the neutral and charged gauge boson of the SM, respectively. The SM differential cross-section of the process with respect to the electron recoil energy $T_{e}$ is given by \footnote{This is presented at the tree level, see Refs. \cite{Erler:2013, Tomalak:2019ibg, Bahcall:1995} for the effect of radiative corrections. We stress that for the low-energy $pp$ and $^7$Be solar neutrinos, the fractional changes due to radiative corrections and to the use of the more accurate weak mixing angle are also less than and of order $1\%$ as discussed in Ref.\cite{Bahcall:1995} and hence they can be safely ignored.} 
\begin{widetext}
	\begin{align} 
		\begin{split}
			\left[\frac{d\sigma_{\nu_\ell}}{dT_{e}}\right]_{\mathrm{SM}} = \frac{G_F^2 m_e}{2\pi} &\Bigg[ (g_V+g_A)^2 + (g_V-g_A)^2 \left(1-\frac{T_{e}}{E_\nu}\right)^2 - (g_V^{2}-g_A^{2}) \frac{m_e T_{e}}{E_\nu^2} \Bigg],
		\end{split}
		\label{eq:sm_nue}
	\end{align}
\end{widetext}
where $m_e$ is the electron mass, $E_\nu$ denotes the initial neutrino energy, and the Fermi coupling constant is $G_F=1.1663787 \times 10^{-5}$ GeV$^{-2}$. 
Furthermore, the neutrino flavour specific vector and axial-vector couplings to electrons respectively are   
\begin{align}
	g_V =& -\frac{1}{2}+2s_W^2 + \delta_{\ell e},\\
	g_A =& -\frac{1}{2}+\delta_{\ell e},
\end{align}
where the abbreviation $s_W=\sin\theta_W$ with $\theta_W$ being weak mixing angle is used, and its value is predicted as $s^2_W = 0.23873$ in the framework the $\overline{\text{MS}}$ scheme at the low-energy regime \cite{ParticleDataGroup:2024cfk}.  Both couplings depend on the flavor $\ell=e,\mu,\tau$ of the incoming neutrino. 
In the above definitions, the Kronecker delta symbol $\delta_{\ell e}$ represents the NC ($\ell=e$) and CC ($\ell=\mu, \tau$) nature of the process. For the $\nu_e$, we have both contributions, while for the $\nu_\mu$ and $\nu_\tau$ cases, only the NC contributes.  

A unique characteristic of neutrino-electron scatterings is that the process is highly directional in nature. This means that the outgoing electron has very small angles with respect to the direction of the incoming neutrino \cite{Formaggio:2012cpf}. This remarkable property has been utilized in many experiments involving neutrinos, especially for detecting solar neutrinos, such as Super-Kamiokande \cite{Kamiokande-II:1989hkh, Super-Kamiokande:2001ljr}, SNO \cite{SNO:2002tuh, SNO:2003bmh} and BOREXINO \cite{Borexino:2008fkj} experiments.
Neutrinos coming from the Sun are able to induce elastic $\nu_e-e$ scattering events in DD experiments of dark matter as the incoming neutrino interacts with the electron cloud. Neutrino-electron scatterings, together with CE$\nu$NS, play an important role as neutrino backgrounds in such experiments. The signal is in principle reducible, but in practice, it is hard to be removed completely in experiments. The neutrino-electron scattering hence has a strong degeneracy with the corresponding DM-electron signals. 

\subsection{New Light Mediator Models}\label{sec:BSM}
At low-energy neutrino experiments, new physics effects can be examined on the elastic neutrino-electron scattering in the presence of light mediators that couple to SM leptons. 
In this work, we consider the so-called universal light mediator models \cite{Cerdeno:2016sfi, Abdallah:2015ter} involving scalar, vector, and tensor interactions allowed by Lorentz invariance, and the anomaly-free $U(1)'$ extensions of the SM with the new vector mediator such as the $L_e-L_\mu$, $L_e-L_\tau$ and $L_\mu-L_\tau$ \cite{He:1991qd, Coloma:2022umy, Gninenko:2020xys} gauge models in which the charges are exclusively leptonic.
The light mediator models are constructed to include only a few new particles and interactions and can be considered as a limit of a more general BSM scenario. 
By construction, this type of model can be described in terms of only a few free parameters such as coupling constants and masses. Such models allow us to research broad classes of new physics signatures without specifying a full high-energy theory.

\begin{figure}[!h]
	\centering
	\includegraphics[scale=0.65]{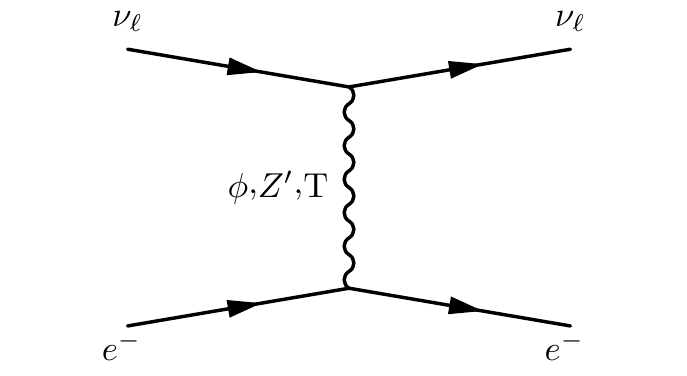}
	\caption{Feynman diagram for neutrino-electron scattering mediated by the new mediators $\phi$, $Z'$, or $\mathrm{T}$.}
	\label{fig:nue_diag_bsm}
\end{figure}
In the presence of the light mediators under consideration, the Feynman diagram for neutrino-electron scattering is shown in Fig. \ref{fig:nue_diag_bsm}. We will calculate the neutrino-electron scattering contributions of scalar, vector, tensor and lepton flavor-dependent of $U(1)'$ mediators with a mass $m_X$ and a defined coupling $g_X = \sqrt{g_X^{e}g_X^{\nu_\ell}}$ (where $g_X^{e}$ and $g_X^{\nu_\ell}$ denote electron and neutrino couplings of $X$ model, respectively),
and investigate how well these can be constrained through the recent data of DD experiments.
Particularly, the low recoil energy data in PandaX-4T may place competitive or stronger bounds on the $m_X$-$g_X$ plane. For a mediator with comparable mass with the typical momentum transfer, its existence can induce changes in the electron recoil spectrum that lead to stronger constraints. 

\subsubsection{Universal scalar}
First we consider a new light scalar mediator $\phi$ coupling
universally to all neutrinos and leptons in the SM.
The relevant interaction Lagrangian can be written as \cite{Cerdeno:2016sfi} %AtzoriCorona:2022moj}
\begin{align}
\begin{split}
\mathcal{L}_\phi \supset -  \phi &\biggl[g_{\phi}^e \bar{e} e + g_{\phi}^{\nu_\ell} \bar{\nu_\ell}_R {\nu_\ell}_L  + h.c. \biggr],
%-\frac{1}{2} m_\phi^2 \phi^2,
\end{split}
\end{align}
where the scalar coupling constant $g_{\phi}^e$ is for electron and $g_{\phi}^{\nu_\ell}$ for neutrino. 
In this model, the contribution to the neutrino-electron scattering reads \cite{Cerdeno:2016sfi}  
\begin{align}
\biggl[\frac{d\sigma_{\nu_\ell}}{dT_{e}}\biggr]_\phi =
\frac{(g_\phi^e g_\phi^{\nu_\ell})^2}{4\pi(m_\phi^2+2m_e T_{e})^2}\left[ \frac{T_e m_e^2}{E_\nu^2}\right]
\label{eq:difcross_scalar}
\end{align}
for a neutrino with flavor $\ell$. This universal scalar interaction does not interfere with the SM. Hence, the contribution from the new scalar mediator is directly added to the SM neutrino-electron cross-section.

\subsubsection{Universal vector}
We consider adding a new light vector mediator $Z'$, which universally couples to the SM electron and neutrinos. The relevant Lagrangian can be expressed as 
\begin{align}
\begin{split}
\mathcal{L}_{Z'} \supset Z'_\mu & \biggl[Q^e_{Z'} g_{Z'}^e \bar{e} \gamma^\mu e + Q^{\nu_\ell}_{Z'} g_{Z'}^{\nu_\ell} \bar{\nu_\ell}_L \gamma^\mu  {\nu_\ell}_L  \biggr],
%+\frac{1}{2} m_{Z'} Z'^{\mu} Z'_\mu
\end{split}
\end{align}
where the vector coupling constants $g_{Z'}^e$ and $g_{Z'}^{\nu_\ell}$ are for electron and neutrino, respectively. The individual vector charges of electrons and neutrinos are denoted by $Q^e_{Z'}$ and $Q^{\nu_\ell}_{Z'}$. We apply these charges for a generalized form of anomaly-free UV-complete models including only the SM particles plus right-handed neutrinos \cite{Allanach:2019}. 

The $Z'$ mediator contribution to the neutrino-electron scattering cross section is obtained by substituting \cite{Ballet:2019}
\begin{align}
	g_V \longrightarrow g_V + \frac{Q^e_{Z'} g_{Z'}^e Q^{\nu_\ell}_{Z'} g_{Z'}^{\nu_\ell}}{2\sqrt{2}G_F(m_{Z'}^2+2m_e T_e)}
	\label{eq:vec_chrg}
\end{align}
into Eq.\eqref{eq:sm_nue}. This expression is valid in both low- and high-mass regimes; it should be noted that typical momentum transfer in DM direct detection experiments is of the order of about 100 keV.
For the universal vector model $Q^e_{Z'}=Q^{\nu_\ell}_{Z'}=1$. These features lead to a similar cross-section formula with the widely studied $B-L$ model in which $Q^e_{Z'}=Q^{\nu_\ell}_{Z'}=-1$. 
The difference between these two models can be observed in the CE$\nu$NS process, where the vector charges of quarks are scaled by a factor of $1/3$ \cite{DeRomeri:2022twg, Demirci:2023tui}. However, in the neutrino-electron scattering process, both models have identical predictions.

\subsubsection{Universal tensor}
We now consider a new light tensor mediator $\mathrm{T}$, which universally couples to the SM electron and neutrinos. The relevant interaction Lagrangian read \cite{Barranco:2011wx}
\begin{align}
\begin{split}
\mathcal{L}_{\mathrm{T}} \supset &\biggl[ g_\mathrm{T}^{e} \bar{e} \sigma^{\mu\nu} e -  g_\mathrm{T}^{\nu_l} \bar{\nu_\ell}_R \sigma^{\mu\nu} {\nu_\ell}_L\biggr] \mathrm{T}_{\mu\nu},
\end{split}
\end{align}
where $\sigma_{\mu\nu} =
i(\gamma_\mu\gamma_\nu-\gamma_\nu\gamma_\mu)/2$. The coupling $g_{\mathrm{T}}^e$ is for electron and $g_{\mathrm{T}}^{\nu_\ell}$ for neutrinos.  This interaction contribution to neutrino-electron scattering is given by \cite{DeRomeri:2022twg}
\begin{align}
\begin{split}
\biggl[\frac{d\sigma_{\nu_\ell}}{dT_{e}}\biggr]_{\mathrm{T}} =
&\frac{(g_\mathrm{T}^e g_\mathrm{T}^{\nu_\ell})^2}{2\pi (m_{\mathrm{T}}^2 + 2m_eT_{e})^2} m_e \Bigg[1 + 2\left(1-\frac{T_{e}}{E_\nu}\right) \\ & + \left(1-\frac{T_{e}}{E_\nu}\right)^2- \frac{m_e T_{e}}{E_\nu^2} \Bigg].
\end{split}
\label{eq:difcross_tensor}
\end{align}

\subsubsection{Lepton flavor-dependent $U(1)'$ models: $L_e-L_\mu$, $L_e-L_\mu$, and $L_\mu-L_\tau$}
The $U(1)'$ models we are interested in are exclusively leptonic. To get a perspective, consider the linear combination \cite{Foot:1992ui}
\begin{align}
	L'= c_e L_e + c_\mu L_\mu + c_\tau L_\tau + c_B B,
\end{align}
where $L_\ell$ is the lepton number with $\ell=e,\mu,\tau$ and $B$ the baryon number with their corresponding coefficients $c_j$ with $j=e,\mu,\tau,B$. It has been assumed that quark charges are universal to avoid unobserved flavor-changing in the quark sector. When the SM is extended to $SU(3)_C \times SU(2)_L \times U(1)_Y \times U(1)'$, we encounter the anomaly constrains $c_e^3 + c_\mu^3 + c_\tau^2 = 0 \quad \text{and} \quad 3c_B + c_e + c_\mu + c_\tau =0$. 
All other gauge anomaly relations are not independent from these equations. The constraints redefine the $L'$, which leads to an additional contribution to the electric charge.
From the mixed gauge-gravitational anomaly cancellation \cite{Alvarez-Gaume:1983ihn}, another further independent relation $c_e+c_\mu+c_\tau=0$ can be included, which implies that $c_B=0$ and either $c_e$, $c_\mu$, or $c_\tau$ is equal to zero. 
The first case gives $L'=L_\mu-L_\tau$, the second $L'=L_e-L_\tau$, while the third $L'=L_e-L_\mu$. 

These are among the simplest models to feature a vector boson $Z'$ since no fermions beyond those already present in the SM, which are required to cancel gauge anomalies \cite{He:1991qd}.
The phenomenology of these gauge groups can be varied from that of other dark photon models. For instance, the considered gauge bosons do not have coupling with the charged $W^\pm$ bosons at tree level. These are also purely coupled to the lepton family number difference of the corresponding charged leptons and neutrinos and not to the baryon number.  
Consequently, new constraints arise from neutrino experiments \cite{Bauer:2018onh}. Studies on these types of mediator have been conducted using stopped-pion neutrino sources \cite{AtzoriCorona:2022moj}, solar neutrino \cite{Coloma:2022umy}, DD \cite{DeRomeri:2024dbv} and collider experiment \cite{Bauer:2018onh}.

\begin{table}[h]
	\caption{Individual vector charges for lepton flavor-dependent $U(1)'$ mediator models.}
    \begin{ruledtabular}
	\begin{tabular}{c c c c c c c} 
%		\hline
%		\hline
		\textbf{Model} &  & $Q_{Z'}^{e/\nu_e}$ & &  $Q_{Z'}^{\mu/\nu_\mu}$ & & $Q_{Z'}^{\tau/\nu_\tau}$ \\
		\hline
%		universal & & $1$ & & $1$ & & $1$ \\
		$L_e-L_\mu$ & & $1$ & & $-1$ & & $0$ \\
		$L_e-L_\tau$ & & $1$ & & $0$ & & $-1$ \\
		$L_\mu-L_\tau$ & & $0$ & & $1$ & & $-1$\\
%		\hline
%		\hline
	\end{tabular}
    \end{ruledtabular}
	\label{tab:modelcharge}
\end{table}
We list the vector charges for each model in Table \ref{tab:modelcharge}. The couplings for new physics contributions are weighted by these vector charges. 
Consequently, $Z'$ contributions from the $L_e-L_\mu$ and $L_e-L_\tau$ models can be easily obtained by implementing Eq.\eqref{eq:vec_chrg} with their gauge charges into Eq. \eqref{eq:sm_nue}.
On the other hand, for the $L_\mu-L_\tau$ model, there is no direct coupling to electrons at tree-level, hence its effect appears at the loop-level. Namely, the $Z'$ contribution of the $L_\mu-L_\tau$ model first arises at one-loop level for $\nu_\mu-e$ and  $\nu_\tau-e$ scatterings  (but vanishes for $\nu_e-e$ scattering \footnote{Contribution to $\nu_e-e$ scattering arises first at two-loop level through $Z^0-Z'$ mixing, hence it is negligible.}). These contributions can be obtained by making the following replacement in Eq.\eqref{eq:sm_nue}  \cite{Altmannshofer:2019}
\begin{align}
	g_V \longrightarrow g_V - \frac{\sqrt{2} \alpha_{\text{em}}g^e_{Z'} g^{\nu_\ell}_{Z'}(\delta_{\ell \mu}-\delta_{\ell \tau})}{\pi G_F (m_{Z'}^2+2m_e T_e)}\epsilon_{\tau \mu}(|\vec{q}|),
	\label{eq:lmlt}
\end{align}
where the $\nu_\mu e$ and  $\nu_\tau e$ scattering contributions differ by a relative minus sign. The factor $\alpha_{\text{em}}$ denotes the fine-structure constant and the $\epsilon_{\tau \mu}$ can be approximated as
\begin{align}
\begin{split}
	\epsilon_{\tau \mu}(|\vec{q}|)&=\int_{0}^{1}x(1-x)\ln\left(\frac{m_\tau^2+x(1-x)|\vec{q}|^2}{m_\mu^2+x(1-x)|\vec{q}|^2}\right)\\
	&\approx \frac{1}{6}\ln\left(\frac{m_\tau^2}{m_\mu^2}\right).
\end{split}
\end{align}

\subsection{Event Rate Spectra} \label{sec:ER}
The differential event rate, which is obtained by multiplying the cross section with neutrino flux, can be written as
\begin{align}
	\left[\frac{dR}{dT_{e}}\right]^i_X = Z_{\text{eff}}(T_e) \int_{E_{\nu}^{\text{min}}}^{E_{\nu,i}^{\text{max}}} dE_\nu \frac{d\Phi^i}{dE_\nu} \left[\frac{d\sigma}{dT_{e}}\right]^{\nu e}_X,
\end{align}
where the minimum neutrino energy is given by
\begin{align}
	E_{\nu}^{\text{min}} = \frac{1}{2}\left(T_e+\sqrt{T_e^{2} + 2m_e T_e} \right),
\end{align}
while we take the kinematic endpoints of each of the neutrino fluxes for the maximum energy $E_{\nu,i}^{\text{max}}$. The minimum energy is necessary to trigger the electron recoil energy.
The differential flux of neutrino is represented by $d\Phi^i(E_\nu)/dE_\nu$, with $i=pp,^7\mathrm{Be}, \text{ etc.}$ referring to the main components of solar neutrino flux. Relevant to neutrino-electron scatterings, we take $pp$ and $^7$Be solar neutrinos in this work.
The index $X=\mathrm{SM}, \phi, Z', \cdots$ indicates different models. 
As the $T_e$ around a few keV comparable to the electron binding energy, the approximation of free-electron is generally no longer valid due to the atomic ionization effect \cite{Chen:2016eab}. To resolve this, we need to multiply the cross-section with the number of effective electron charges that can be ionized, represented by $Z_{\mathrm{eff}}(T_e)=\sum_{\alpha}n_\alpha\theta(T_e - B_\alpha)$. 
Here, $n_\alpha$ and $B_\alpha$ are the number of electrons and their binding energy in the atomic shell $\alpha$, respectively, while $\theta(x)$ represents a Heaviside step function. We present this factor in Table \ref{tab:XeEff} for xenon target \cite{xraydata:2009}. This is necessary to rectify the cross-section derived under the free electron approximation hypothesis, in which electrons are considered to be free and at rest \cite{Kouzakov:2017hbc, Hsieh:2019hug}.
\begin{table}[hbt]
	\caption{Number of effective electron charges for xenon as a function of energy deposition $T_{e}$ \cite{xraydata:2009}.} \label{tab:XeEff}
    \begin{ruledtabular}
	\begin{tabular}{cc} 
%		\hline
		\textbf{$Z_{\mathrm{eff}}(T_{e})$} &  $T_{e}$ in eV\\
		\hline
		$54$ & $T_{e}>34561$ \\
		$52$ & $34561\geq T_{e}>5452.8$ \\
		$50$ & $5452.8\geq T_{e}>5103.7$ \\
		$48$ & $5103.7\geq T_{e}>4782.2$ \\
		$44$ & $4782.2\geq T_{e}>1148.7$ \\
		$42$ & $1148.7\geq T_{e}>1002.1$ \\
		$40$ & $1002.1\geq T_{e}>940.6$ \\
		$36$ & $940.6\geq T_{e}>689.0$ \\
		$32$ & $689.0\geq T_{e}>676.4$ \\
		$26$ & $676.4\geq T_{e}>213.2$ \\
		$24$ & $213.2\geq T_{e}>146.7$ \\
		$22$ & $146.7\geq T_{e}>145.5$ \\
		$18$ & $145.5\geq T_{e}>69.5$ \\
		$14$ & $69.5\geq T_{e}>67.5$ \\
		$10$ & $67.5\geq T_{e}>23.3$ \\
		$4$ & $23.3\geq T_{e}>13.4$ \\
		$2$ & $13.4\geq T_{e}>12.1$ \\
		$0$ & $12.1\leq T_{e}$ \\
	%	\hline
	\end{tabular}
\end{ruledtabular}
\end{table}

Neutrinos undergo oscillations during their propagation from the Sun to Earth. Solar neutrinos reach a detector as a mixture of $\nu_e$, $\nu_\mu$, and $\nu_\tau$. 
To take this oscillation into consideration, the cross-section hence needs to be weighted with the relevant survival probabilities
\begin{align}
	\left[\frac{d\sigma}{dT_{e}}\right]^{\nu e}_X = P_{ee} \left[\frac{d\sigma_{\nu_e}}{dT_{e}}\right]_X + \sum_{f=\mu,\tau} P_{ef} \left[\frac{d\sigma_{\nu_f}}{dT_{e}}\right]_X,
\end{align}
where the conversion probabilities of $\nu_e$ to $\nu_\mu$ and $\nu_\tau$ are given by $P_{e\mu}=(1-P_{ee})\cos^2\vartheta_{23}$ and $P_{e\tau}=(1-P_{ee})\sin^2\vartheta_{23}$, respectively. The $P_{ee}$ denotes the survival probability of $\nu_e$ which satisfies \cite{Maltoni:2015kca}
\begin{align}
	\begin{split}
		P_{ee} = & \cos^2 (\vartheta_{13}) {\cos^2 (\vartheta_{13}^m)}\left( \frac{1}{2} + \frac{1}{2} \cos (2\vartheta_{12}^m) \cos (2\vartheta_{12}) \right) \\ &+ \sin^2 (\vartheta_{13}) {\sin^2(\vartheta_{13}^m)}.
		\label{Pee}
	\end{split}
\end{align}
where the label $m$ represents the matter effect. The survival probabilities depend on neutrino mixing angles of $\vartheta_{12}, \vartheta_{13}$ and $ \vartheta_{23}$.
We consider the day-night asymmetry due to the Earth matter effect in the calculation of these probabilities. We take the normal-ordering neutrino oscillation parameters from the latest 3-$\nu$ oscillation of NuFit-5.3, without the Super-Kamiokande atmospheric data \cite{Esteban:2020cvm}. 

The predicted number of elastic neutrino-electron scattering events is calculated from
%\begin{widetext}
\begin{align}
\begin{split}
	R^k_X=&\varepsilon N_{t}\int_{T_e^k}^{T_e^{k+1}}dT_e \text{ } \mathcal{A}(T_e) \int_{0}^{T_e^{'\text{max}}}dT_e' \text{ } \mathcal{R}(T_e,T_e') \\ &\times \sum_{i=pp,^7\text{Be}} \left[\frac{dR}{dT'_{e}}\right]^i_X,
\end{split}
\end{align}
%\end{widetext}
where the factor $N_{t}$
denotes the number of target nuclei per unit mass of the detector material. 
The reconstructed and true electron recoil energies are given by $T_e$ and $T_e'$, respectively. The factors $\mathcal{A}(T_e)$ and $\mathcal{R}(T_e,T_e')$ are detector efficiency and smearing functions, respectively, which should be taken into account to accurately simulate the signal.
We take the PandaX-4T experimental efficiency from Ref. \cite{PandaX:2024cic} and the normalized Gaussian smearing function with energy resolution $\sigma =  0.073+0.173 T_e - 0.0065 T_e^2  + 0.00011 T_e^3$ from Ref. \cite{PandaX:2022ood}. 
To obtain the reported count of the experiment, this rate is also multiplied by the exposure factor $\varepsilon$. 
The Run0 data of the PandaX-4T experiment has exposure of $198.9\text{ ton} \cdot \text{day}$ and Run1 data has $363.3\text{ ton} \cdot \text{day}$. Meanwhile, the maximum recoil energy satisfies
\begin{align}
	T_e^{'\text{max}} = \frac{2E_\nu^2}{2E_\nu + m_e}.
\end{align}
This relation explains that lighter targets improve the maximum recoil energy produced in the detector.

\section{Data Analysis Details}\label{sec:stat}
We now discuss the implementation of the statistical analysis with experimental data. 
We analyze recent low-energy electron recoil data from the PandaX-4T experiment \cite{PandaX:2024cic}, followed the unblinding of Run0 and Run1 data in Ref. \cite{PandaX:2024zbo}.
The experiment is part of the China Jinping Underground Laboratory (CJPL). It uses a time projection chamber with $3.7$ tonnes of liquid xenon (LXe) to measure the precise position, timing, and energy of each event \cite{PandaX:2018wtu}. Since it started operating, the experiment has been conducting some new physics searches such as WIMP, axion particle, neutrino magnetic moment, and dark photon. The collaboration has recently reported two data sets with electron recoil $<30$ keV. The first data of its commissioning run is termed Run0 and operated from 28 November 2020 to 16 April 2021. A dedicated offline xenon distillation was carried out to reduce the tritium background identified after Run0 \cite{Cui:2020bwf}. The experiment resumed operations, collecting data from 16 November 2021 to 15 May 2022 which is termed Run1. A description of the signal response model employed in the PandaX-4T analysis results can be found in Ref. \cite{PandaX:2024med}. As one of the background sources in the DD experiment, the neutrino-electron scattering contribution in the SM is adequately known and flat with respect to the recoil energy and thus is usually subtracted in standard dark matter analyses \cite{AtzoriCorona:2022jeb}. Nevertheless, the appearance of certain BSM scenarios can significantly enhance the observation signal.
In this regard, we investigate the contribution of light mediators to the neutrino-electron scattering background. Their effect is analyzed in the region of interest (ROI) with low electron recoil data in the experiment under consideration.

For statistical analysis of the new physics parameter(s) of interest $\mathcal{S}$, we use a Gaussian $\chi^2$ function \cite{Baker:1983tu, Fogli:2002pt} 
\begin{widetext}
	\begin{align}
		\begin{split}
		\chi^2(\mathcal{S}) = \mathrm{min}_{(\alpha_i,\beta_i)} \biggl[ \sum_{k=1}^{30} &\Bigg( \frac{ R_\text{exp}^{k}(\mathcal{S}; \alpha,\beta) - R_{obs}^{k} }{\sigma^{k}}\Bigg)^2  + \sum_{i}\left(\frac{\alpha_i}{\sigma_{\alpha_i}}\right)^2 +  \sum_{i}\left(\frac{\beta_i}{\sigma_{\beta_i}}\right)^2 \biggr],
		\end{split} \label{eq:chi2}
	\end{align} 
\end{widetext}
where the $R_{\text{obs}}^k$ and $R_{\text{exp}}^k$ denote the observed and expected event rates in the $k$-th energy bin. The expected one consists of SM $R^k_{\text{SM}}$ plus BSM $R^k_{\text{BSM}}$ ($\phi, Z', \cdots$) contribution and other background component $R_{\text{Bkg}}$. 
Explicitly, $R_{\text{exp}}^k(\mathcal{S};\alpha, \beta)=(1+\alpha)(R^k_{\text{SM}}+R^k_{\text{BSM}}(\mathcal{S})) + (1+\beta)R_{\text{Bkg}}^k$ where the nuisance parameters $\alpha$ and $\beta$ account for the uncertainty on the neutrino flux and background normalization. We utilize the background reported by the collaboration in Ref. \cite{PandaX:2024cic} and retain only one nuisance parameter $\beta$ to account for background uncertainty\footnote{We have checked that the obtained results are essentially unaltered with one nuisance parameter to account for background fluctuation.}.
The experimental uncertainty in the $k$-th energy bin is denoted by $\sigma^k$ \cite{PandaX:2024cic}.
The uncertainty of solar neutrino flux is represented by $\sigma_{\alpha}$ and for the background by $\sigma_{\beta}$ where their values are taken from Table 6 of Ref.\cite{Vinyoles:2016djt} and extracted from Table 2 of Ref.\cite{PandaX:2024cic}, respectively.
The SM plus new physics is calculated by implementing solar neutrino fluxes of the Bahcall's spectrum \cite{Bahcall:1989ks} and following normalization of the B16-GS98 standard solar model \cite{Vinyoles:2016djt}. 
We note that the total neutrino-electron event rates we obtained with the above set by summing over all the energy bins are 41.65 for Run0 and 74.68 for Run1, which are in good agreement with the results of PandaX-4T \cite{PandaX:2024cic}.

\begin{figure*}[htb]
	\centering
	\includegraphics[scale=0.40]{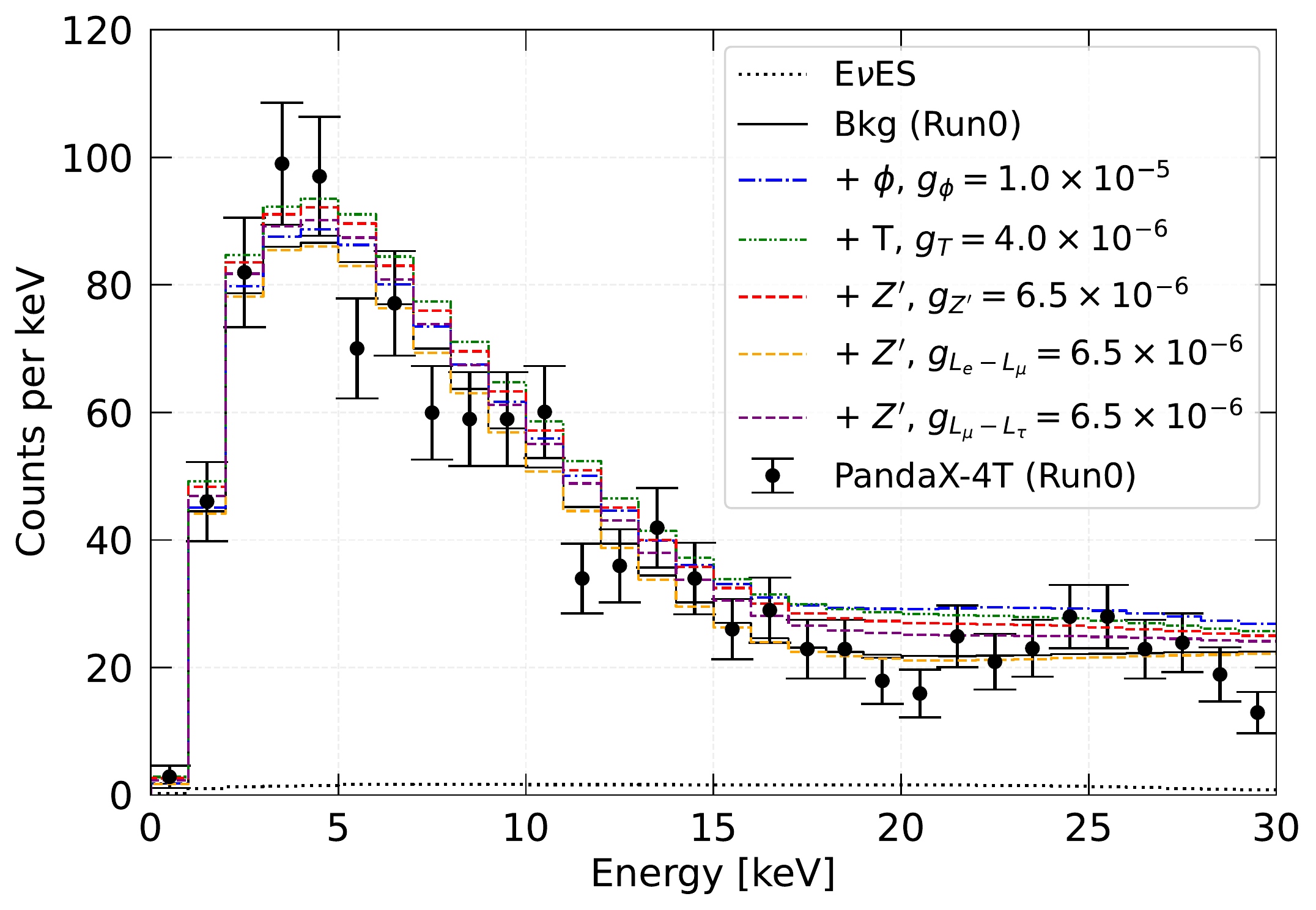}
	\includegraphics[scale=0.40]{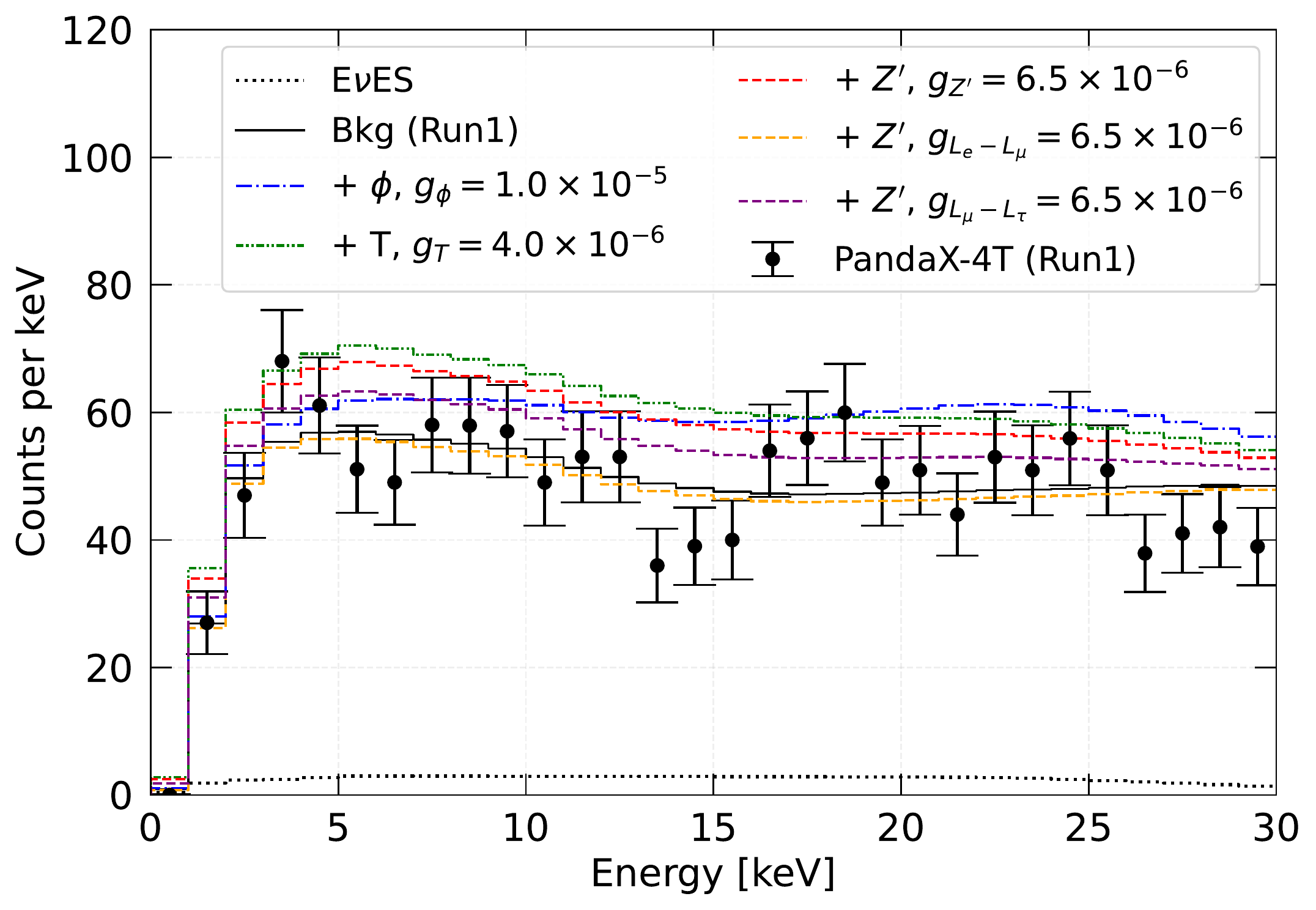}
	\\
	\hspace{7mm} (a) \hspace{82mm} (b)
	\caption{Predicted signals of light mediator models with $m_X=1$ MeV (colored histograms) and experimental data (black points with error bars) as a function of electron recoil energy for (a) Run0 and (b) Run1 datasets of PandaX-4T.  
    The elastic neutrino-electron scattering, denoted by E$\nu$ES, is represented by the dashed black line and the total background by the solid black line.}
	\label{fig:cevns_bsm}
\end{figure*}
\section{Results and Discussion}\label{sec:resdis}
In this section, we provide numerical results obtained from statistical analysis. 
We compare the PandaX-4T low electron recoil data with the expected signal for various light mediator scenarios in Figs. \ref{fig:cevns_bsm}(a) for Run0 and \ref{fig:cevns_bsm}(b) for Run1. They are normalized in $\text{ keV}^{-1}$. We simulate the neutrino-electron scattering signal considering only the $pp$ and $^7$Be solar flux components, which have the main contributions. The predicted energy spectrum has been subtracted from the total background. 
For illustrative purposes, we have set the mediator masses to be $1$ MeV and the coupling constants $g_\phi=1.0\times 10^{-5}$, $g_T=4.0\times 10^{-6}$ and $g_{Z'}=g_{L_e-L_\mu}=g_{L_\mu-L_\tau}=6.5\times 10^{-6}$. We have considered the individual contribution of each model by embedding them into the neutrino-electron scattering signal.
From these plots, we notice that the new physics with the chosen benchmarks may affect the reported background. Having chosen the same values for the universal vector and the $L_e-L_{\mu}$ model, we can see that the latter provides fewer counts than the universal vector case. This is anticipated from the solar neutrino survival probabilities. The spectrum of $L_e-L_{\tau}$ model is not presented in Fig. \ref{fig:cevns_bsm} since it follows the same trend with $L_e-L_{\mu}$. For the $L_\mu-L_\tau$ model, more excess is expected than in the other two flavor-dependent models. 
These types of light mediator are then a general way to improve the measurement of low-energy events.

%%
%%%%
\begin{figure*}[htb]
	\centering
	\includegraphics[scale=0.45]{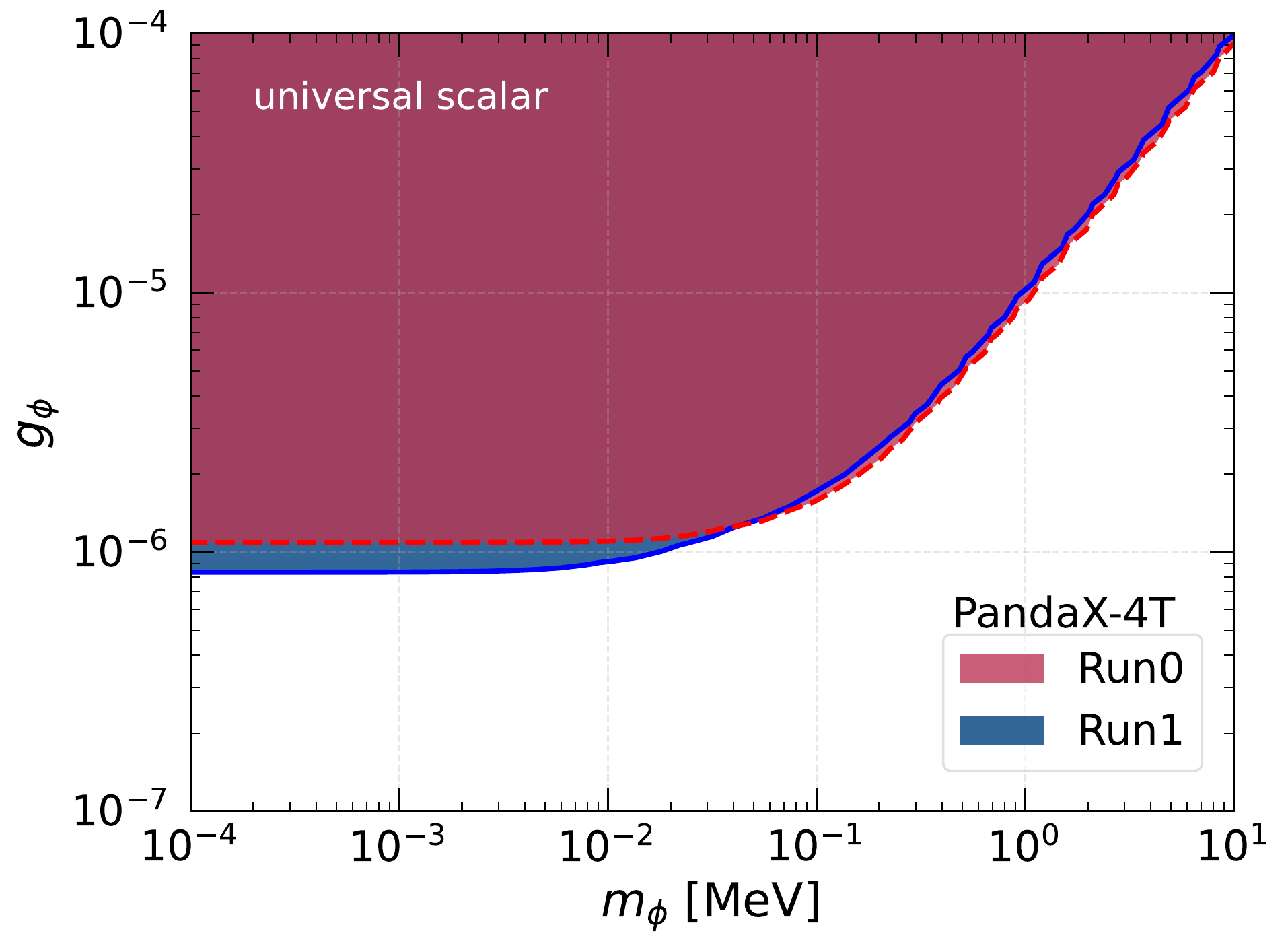}
	\includegraphics[scale=0.45]{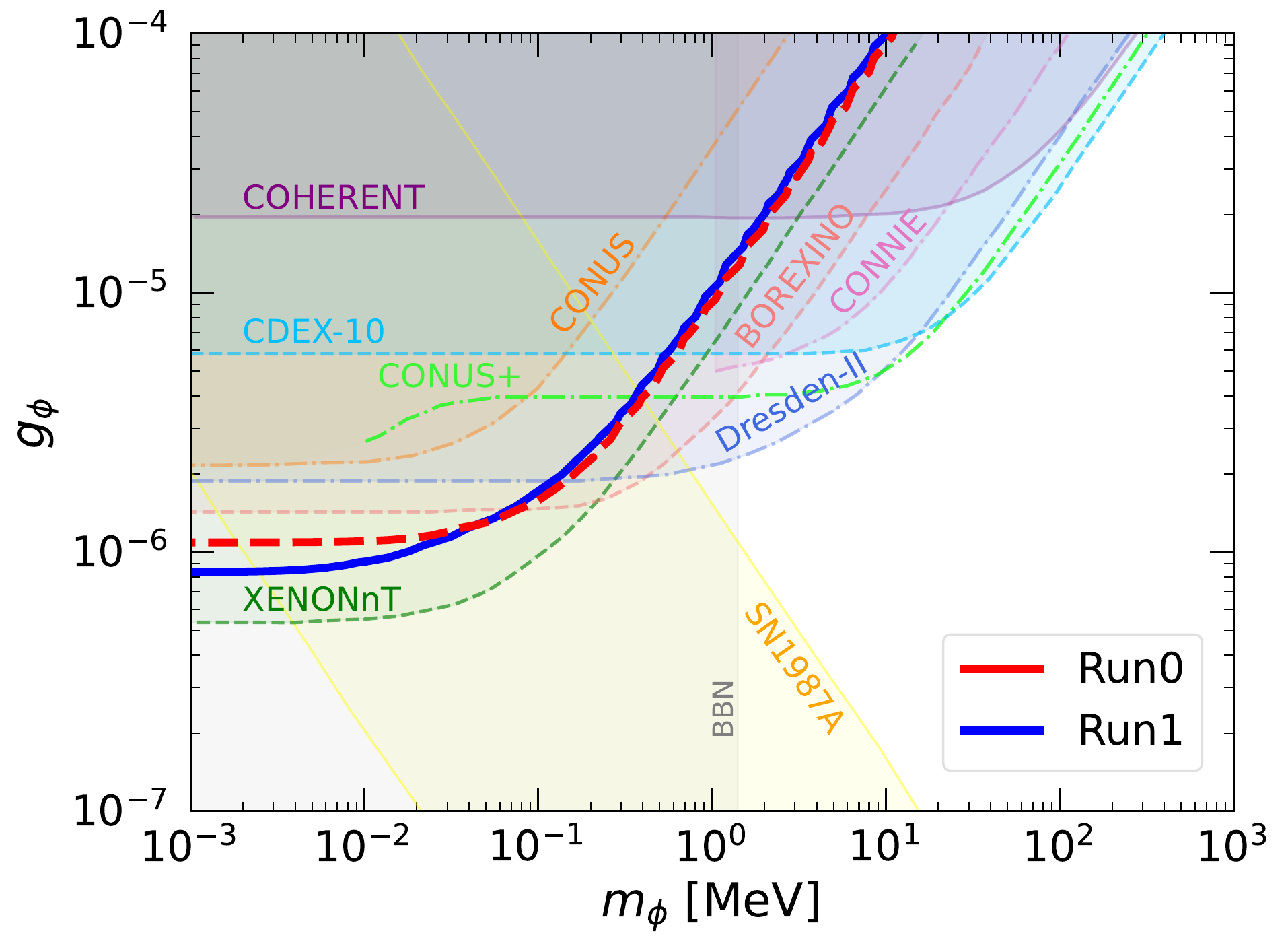}
	\\
\hspace{10mm} (a) \hspace{83mm} (b)
	\\
	%\vspace{-2mm}
	\caption{(a) 90\% C.L. with 2 d.o.f. exclusion regions on the mass-coupling plane of the universal light scalar mediator from the Run0 and Run1 datasets of PandaX-4T. (b) Comparison with existing limits obtained from previous works for COHERENT  \cite{DeRomeri:2022twg}, CONUS \cite{CONUS:2021dwh}, CONUS+ \cite{Chattaraj:2025}, %DUNE \cite{Melas:2023olz}, 
    CONNIE \cite{CONNIE:2019xid}, DRESDEN-II (Fef) \cite{Coloma:2022avw}, BOREXINO \cite{Coloma:2022umy}, XENONnT \cite{A:2022acy} and CDEX-10 \cite{Demirci:2023tui}. Astrophysical limits from Big Bang Nucleosynthesis (BBN) \cite{Blinov:2019gcj} and supernova (SN1987A) \cite{Heurtier:2017} are also shown.
	}
	\label{fig:analysis_s}
\end{figure*}
%%%%%%%
We provide the exclusion regions on the coupling-mass plane for each light mediator model under consideration. We derive the $90 \%$ C.L. upper-limits from 2 degrees of freedom (d.o.f.) analysis of recent low electron recoil datasets (Run0 and Run1) of PandaX-4T experiment by using the $\chi^2$ function \eqref{eq:chi2}. For comparison, we superimpose our results with available constraints derived from various experimental probes such as a stopped-pion source (COHERENT  \cite{Melas:2023olz,DeRomeri:2022twg,Coloma:2022avw, Coloma:2022umy}, CONUS \cite{CONUS:2021dwh,Coloma:2022umy}), nuclear reactors (TEXONO \cite{Bilmis:2015lja}, DRESDEN-II (considering Fef quenching factor) \cite{Coloma:2022avw}, CONNIE \cite{CONNIE:2019xid}, CONUS+ \cite{Chattaraj:2025}), solar neutrino (BOREXINO \cite{Coloma:2022umy,Gninenko:2020xys}), other neutrino-electron scattering (GEMMA \cite{Lindner:2020kko}, CHARM-II \cite{Bilmis:2015lja}, LSND \cite{Bilmis:2015lja}) and other dark matter experiments (NA64 \cite{NA64:2022yly}, XENONnT \cite{Melas:2023olz, A:2022acy}, CDEX-10 \cite{Demirci:2023tui}). Furthermore, we include various collider limits (Mainz \cite{A1:2011yso}, KLOE \cite{ALICE:2012aqc}, BaBar \cite{BaBar:2014zli}, PHENIX \cite{PHENIX:2014duq}, NA48/2 \cite{NA482:2015wmo}), oscillation data limit \cite{Coloma:2022umy, Coloma:2020gfv}, as well as the allowed bound of $(g-2)_\mu$ \cite{Muong-2:2023cdq}. 
Finally, for complementary, we also present astrophysical limits (Big Bang Nucleosynthesis (BBN) \cite{Blinov:2019gcj,Huang:2018}, supernova (SN1987A) \cite{Heurtier:2017,Croon:2020lrf}, cosmology (H0) \cite{Escudero:2019gzq}, and stellar cooling \cite{LiXu:2023}), although their precise prediction will eventually depend on the specific model and thermal history in the early Universe.

\begin{figure*}[htb]
	\centering
	\includegraphics[scale=0.45]{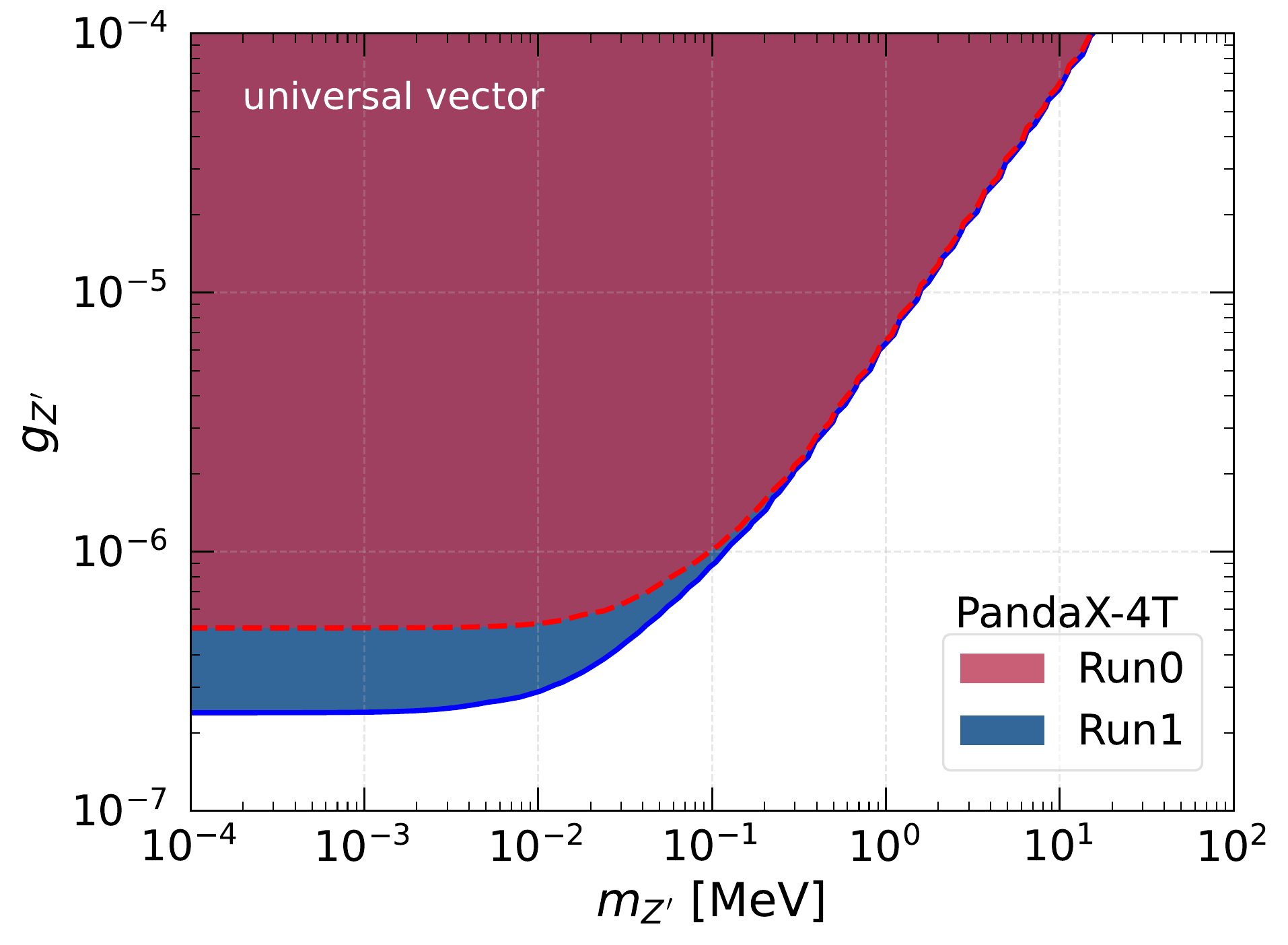}
	\includegraphics[scale=0.45]{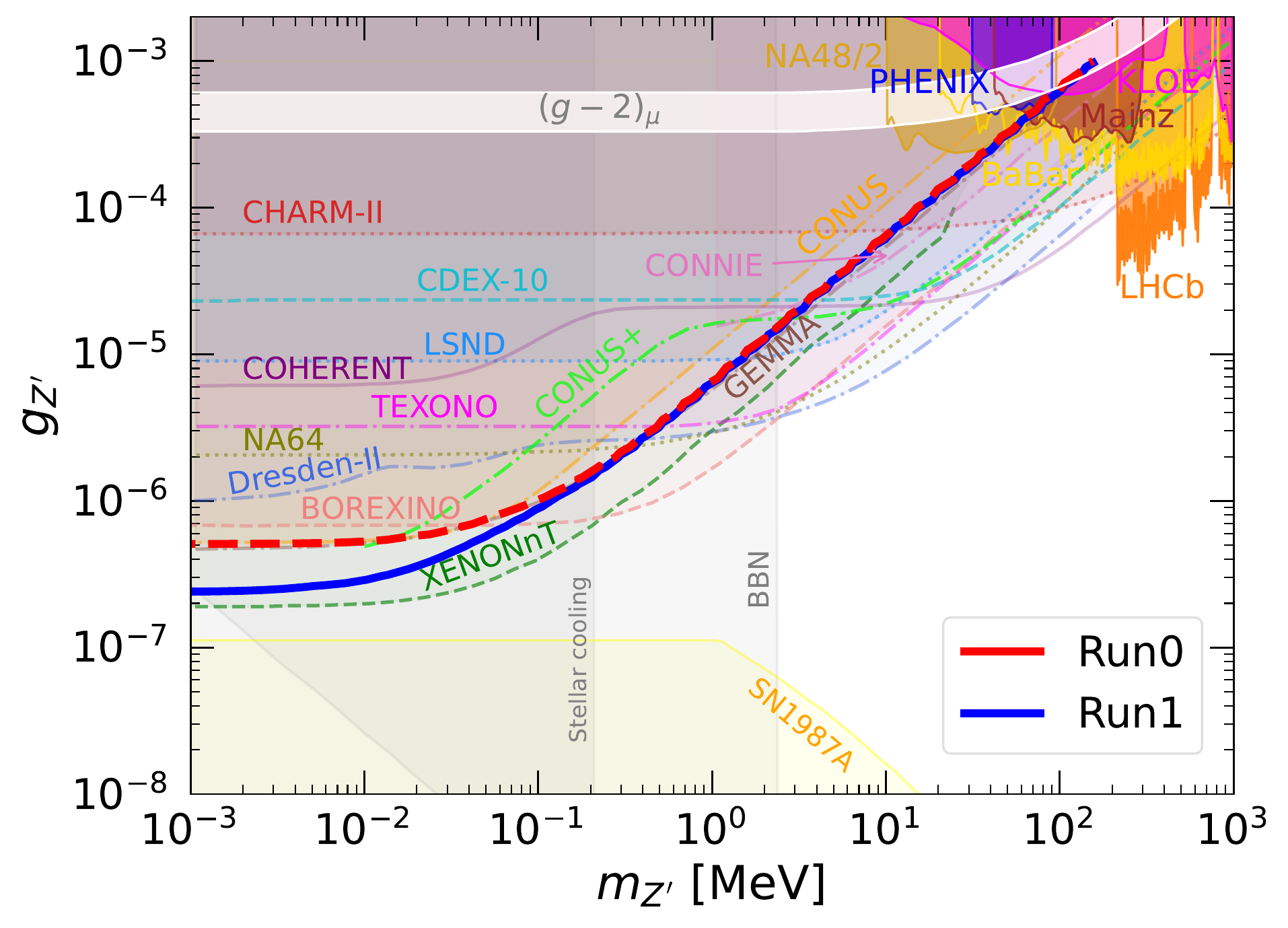}
	\\
\hspace{10mm} (a) \hspace{83mm} (b)
	\\
	%\vspace{-2mm}
	\caption{(a) 90\% C.L. with 2 d.o.f. exclusion regions on the mass-coupling plane of the universal light vector mediator from the Run0 and Run1 datasets of PandaX-4T. (b) Comparison with existing limits obtained from previous works for BOREXINO \cite{Coloma:2022umy}, COHERENT  \cite{DeRomeri:2022twg}, CONUS \cite{CONUS:2021dwh}, CONNIE \cite{CONNIE:2019xid}, CONUS+ \cite{Chattaraj:2025}, TEXONO \cite{Bilmis:2015lja}, DRESDEN-II (Fef) \cite{Coloma:2022avw},  GEMMA \cite{Lindner:2020kko}, CHARM-II \cite{Bilmis:2015lja}, LSND \cite{Bilmis:2015lja}, NA64 \cite{NA64:2022yly}, XENONnT \cite{A:2022acy} and CDEX-10 \cite{Demirci:2023tui}. The limits from colliders (Mainz \cite{A1:2011yso}, KLOE \cite{ALICE:2012aqc}, BaBar \cite{BaBar:2014zli}, PHENIX \cite{PHENIX:2014duq}, NA48/2 \cite{NA482:2015wmo}) and astrophysical
limits (BBN ($\Delta N_{\text{eff}}\simeq 1)$ \cite{Huang:2018}, stellar cooling \cite{LiXu:2023}, supernova (SN1987A) \cite{Croon:2020lrf}) as well as the allowed bound of $(g-2)_\mu$ \cite{Muong-2:2023cdq} are also shown.}
	\label{fig:analysis_v}
\end{figure*}
We show our results in Fig. \ref{fig:analysis_s}(a) on the coupling-mass plane of the universal scalar mediator derived from the PandaX-4T Run0 (red region with dashed line) and Run1 (blue region with solid line) datasets. For $m_\phi \leq 1$ keV, the limit of $g_\phi$ reaches a value of $1.1\times 10^{-6}$ and $8.3\times 10^{-7}$ for Run0 and Run1, respectively. The Run1 data provides a more stringent limit, about 1.3 times, than the Run0 data, as expected from the larger exposure. Next, we overlaid the limits of previous studies outlined above in Fig. \ref{fig:analysis_s}(b). It can be seen that the Run0 and Run1 results provide a more stringent constraint than those obtained from COHERENT (combined CsI+LAr analysis), CDEX-10 (CE$\nu$NS channel), BOREXINO, CONNIE (95\% C.L.), CONUS+ (CE$\nu$NS+ $\nu_e e$ scattering) and Dresden-II in the low-mass region, as well as the limit derived from CONUS ($\nu_e e$ scattering data) in the all-mass region.  
On the other hand, the Run0 and Run1 limits are of the same order of magnitude as the XENONnT limit yet provide no improvement. Furthermore, the current study is complementary to the BBN limit for $m_\phi > 1.4$ MeV.

\begin{figure*}[htb]
	\centering
	\includegraphics[scale=0.45]{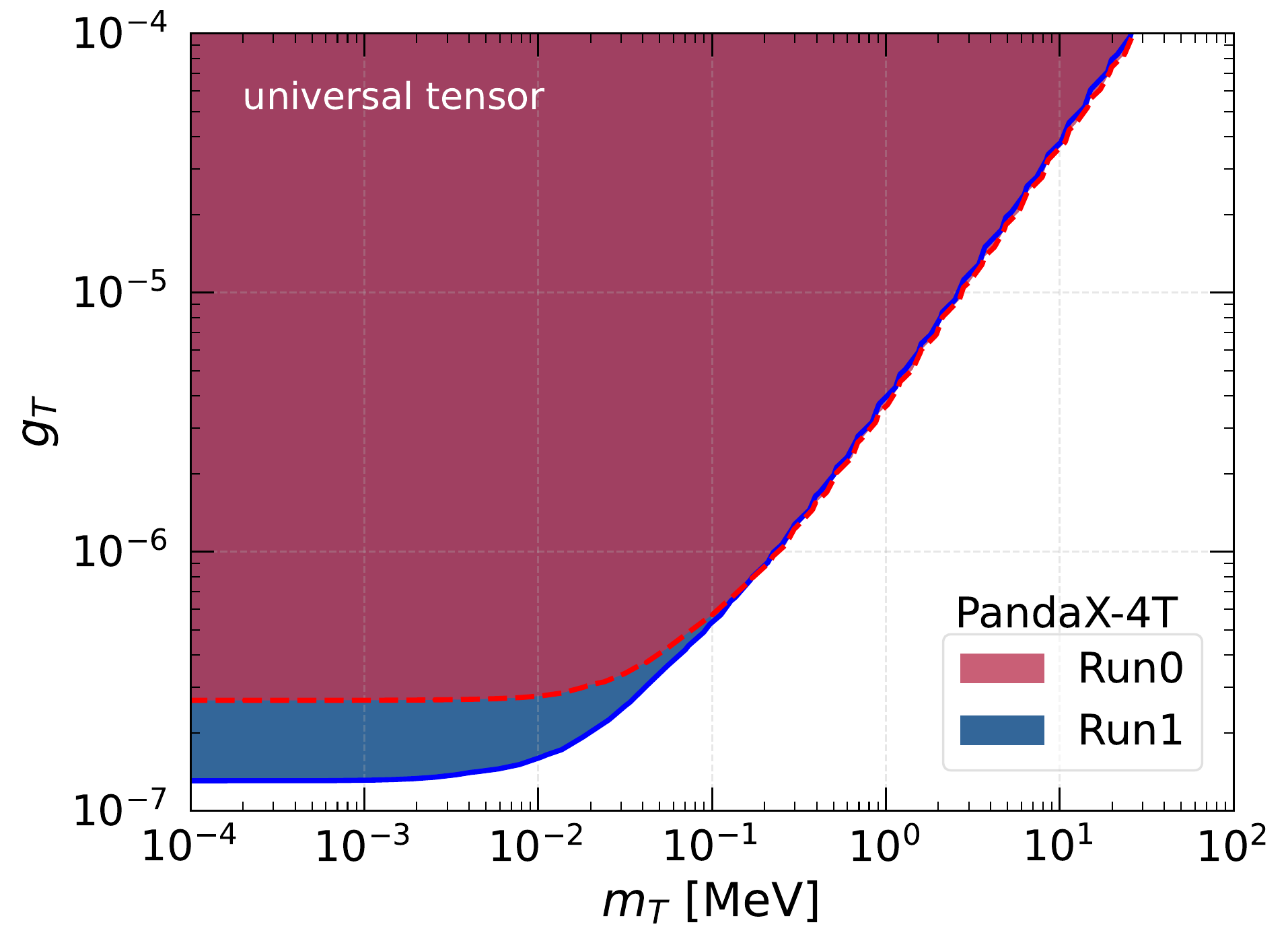}
	\includegraphics[scale=0.45]{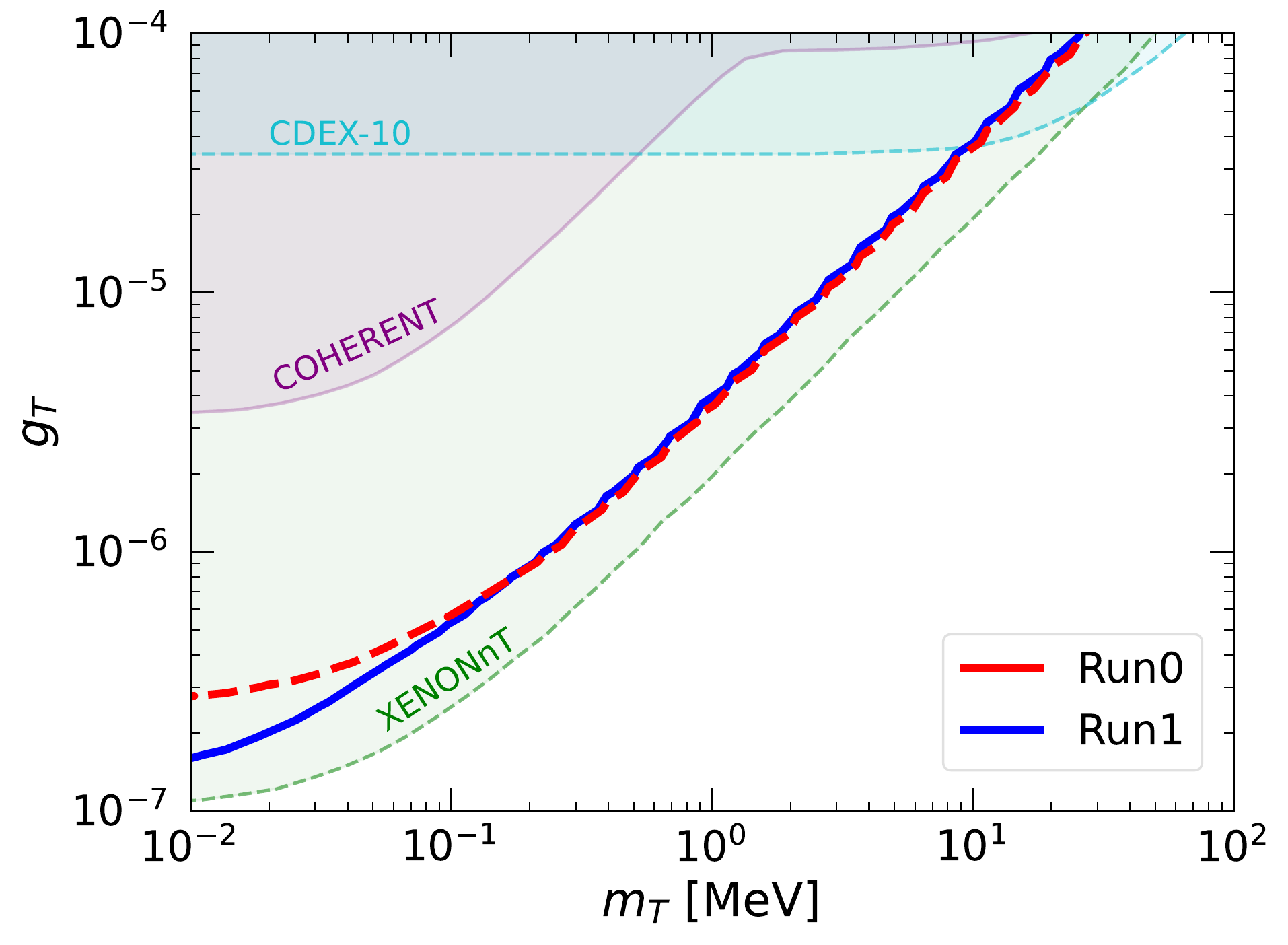}
	\\
\hspace{10mm} (a) \hspace{83mm} (b)
	\\
	%\vspace{-2mm}
	\caption{(a) 90\% C.L. with 2 d.o.f. exclusion regions on the mass-coupling plane of the universal light tensor mediator from the Run0 and Run1 datasets of PandaX-4T. (b) Comparison with existing limits obtained from previous works for COHERENT  \cite{DeRomeri:2022twg}, XENONnT \cite{A:2022acy} and CDEX-10 \cite{Demirci:2023tui}. %DUNE \cite{Melas:2023olz}.
	}
	\label{fig:analysis_t}
\end{figure*}
%%%%
\begin{figure*}[htb]
	\centering
	\includegraphics[scale=0.45]{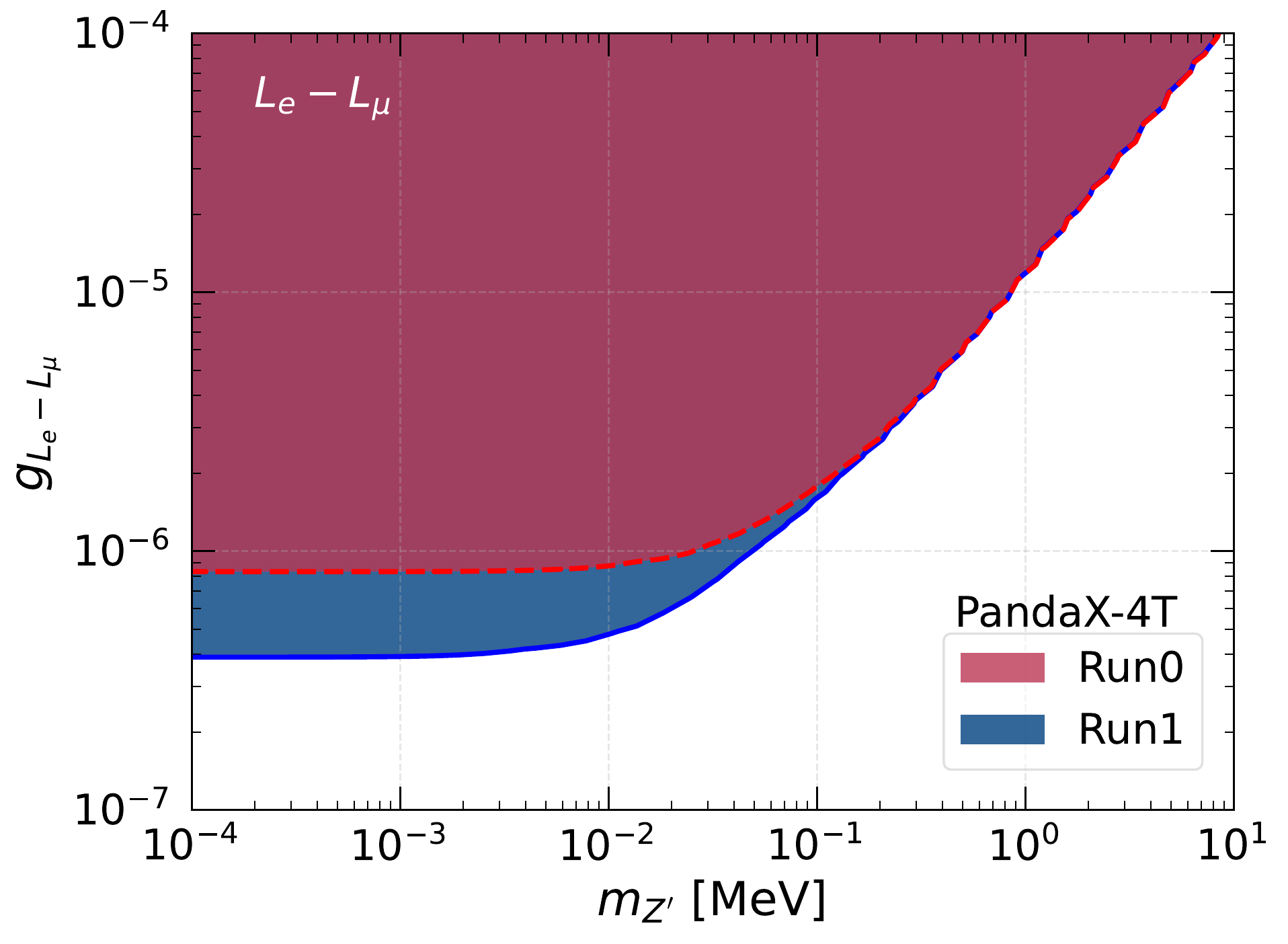}
	\includegraphics[scale=0.45]{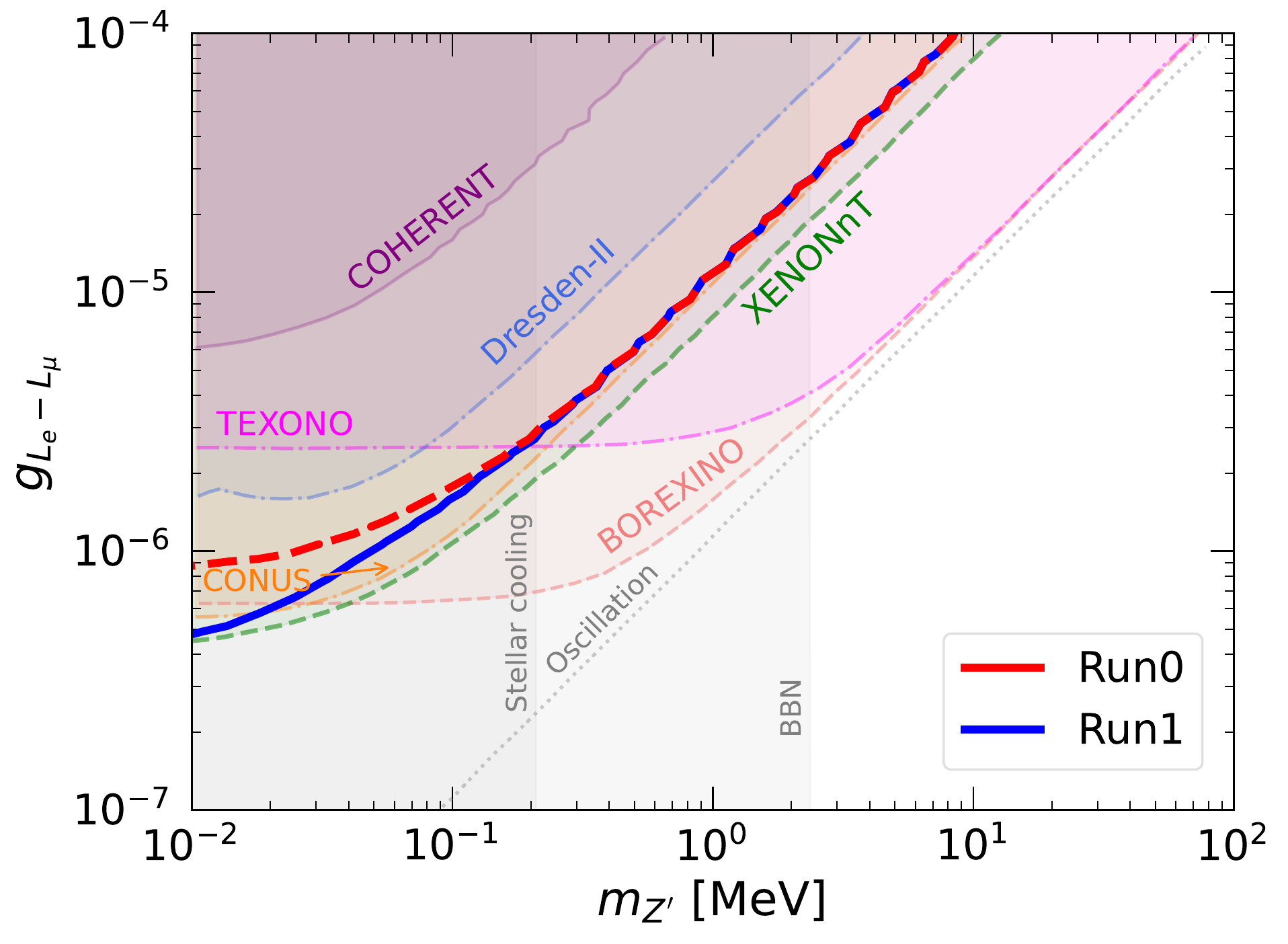}
	\\
\hspace{10mm} (a) \hspace{83mm} (b)
	\\
	\caption{(a) 90\% C.L. with 2 d.o.f. exclusion regions on the mass-coupling plane of the $L_e-L_\mu$ model from the Run0 and Run1 datasets of PandaX-4T. (b) Comparison with existing limits obtained from previous works for COHERENT \cite{Coloma:2022avw, Coloma:2022umy}, CONUS \cite{Coloma:2022umy}, TEXONO \cite{Coloma:2022umy, Coloma:2020gfv}, DRESDEN-II \cite{Coloma:2022avw, Coloma:2022umy}, BOREXINO \cite{Coloma:2022umy}, and XENONnT derived in this work.
    The limits obtained from a global fit to oscillation data \cite{Coloma:2022umy, Coloma:2020gfv},  BBN ($\Delta N_{\text{eff}}\simeq 1)$ \cite{Huang:2018} and stellar cooling \cite{LiXu:2023} are also shown.
	}
	\label{fig:analysis_lelm}
\end{figure*}
\begin{figure*}[htb]
	\centering
	\includegraphics[scale=0.45]{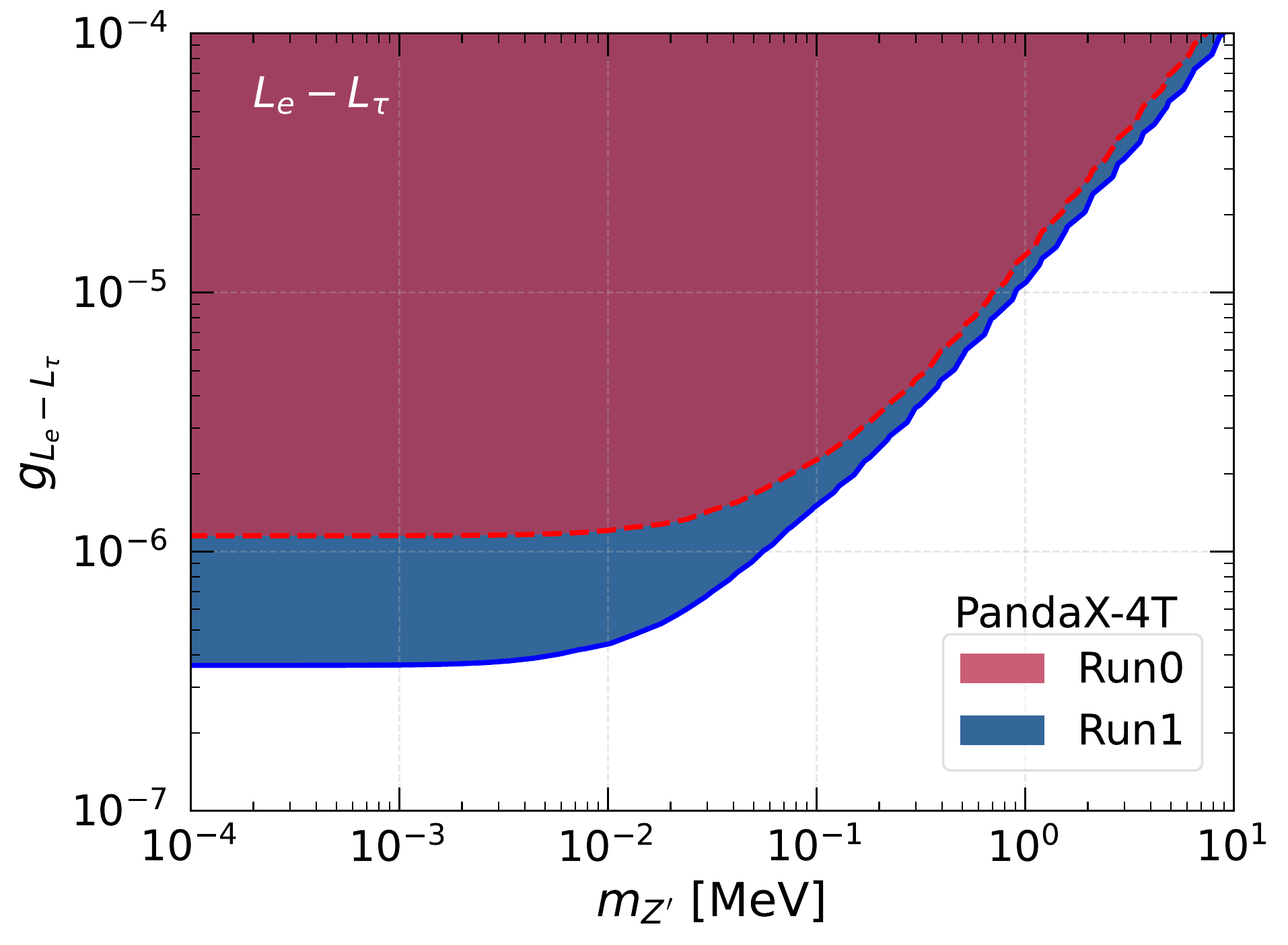}
	\includegraphics[scale=0.45]{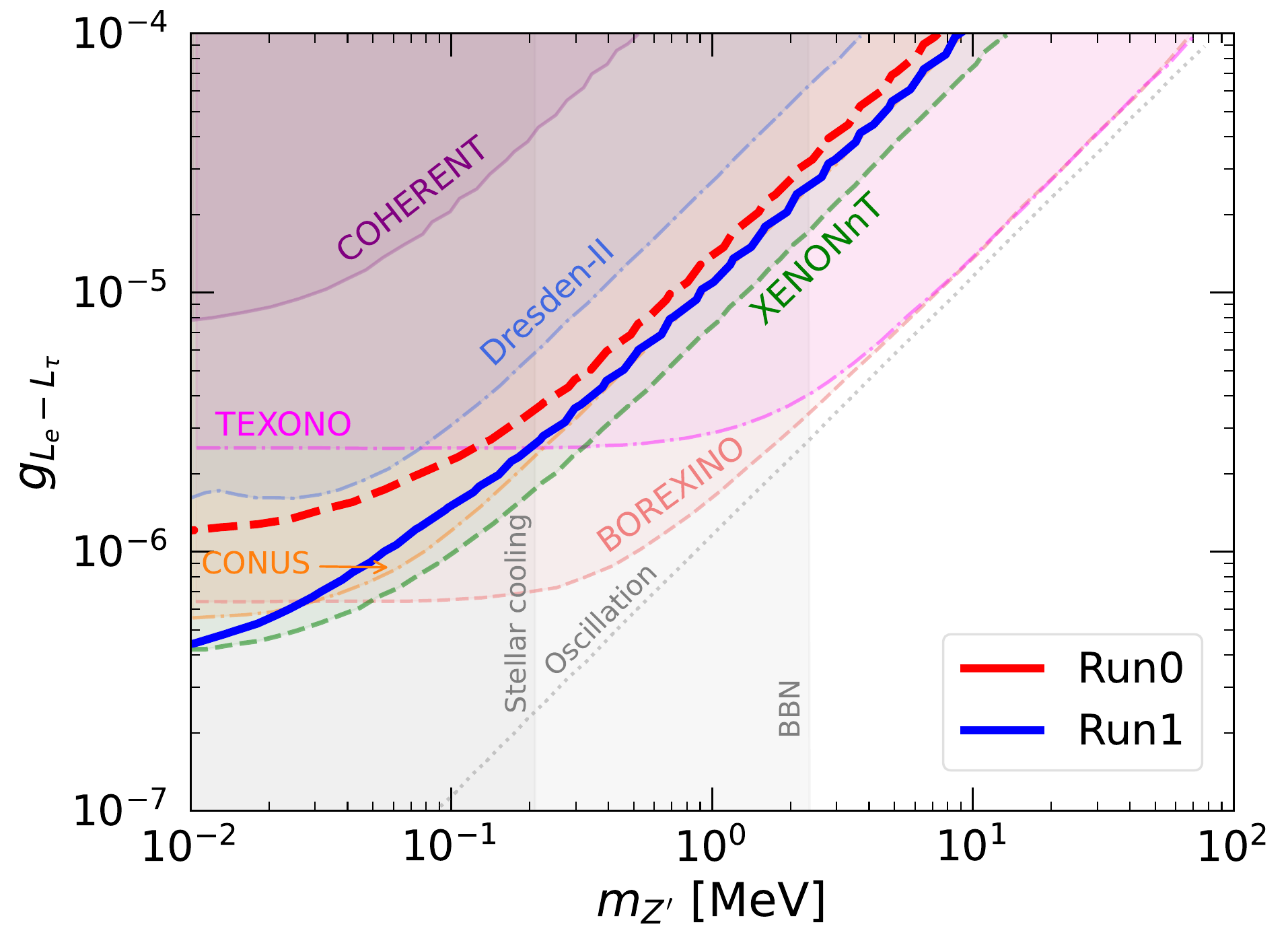}
	\\
\hspace{10mm} (a) \hspace{83mm} (b)
	\\
	\caption{(a) 90\% C.L. with 2 d.o.f. exclusion regions on the mass-coupling plane of the $L_e-L_\tau$ model from the Run0 and Run1 datasets of PandaX-4T. (b) Comparison with existing limits of COHERENT \cite{Coloma:2022avw, Coloma:2022umy}, CONUS \cite{Coloma:2022umy}, TEXONO \cite{Coloma:2022umy, Coloma:2020gfv}, DRESDEN-II \cite{Coloma:2022avw, Coloma:2022umy}, BOREXINO \cite{Coloma:2022umy}, and XENONnT derived in this work.
    The limits obtained from a global fit to oscillation data \cite{Coloma:2022umy, Coloma:2020gfv},  BBN ($\Delta N_{\text{eff}}\simeq 1)$ \cite{Huang:2018} and stellar cooling \cite{LiXu:2023} are also shown.
	}
	\label{fig:analysis_lelt}
\end{figure*}

In Fig. \ref{fig:analysis_v}(a), we show the upper limits with 90\% C.L. (2 d.o.f.) on the coupling-mass plane of the universal vector mediator derived from Run0 and Run1 datasets of PandaX-4T. It can be seen that the upper-limit of $g_{Z'}$ reaches a value of $5.1\times 10^{-7}$ for the Run0 data and $2.4\times 10^{-7}$ for the Run1 data in the region of $m_{Z'} \leq 1$ keV. The Run1 data provides an improvement by about 2.1 times on the limit of $g_{Z'}$ over the Run0 data for the universal vector mediator model in the low-mass region.

In Fig. \ref{fig:analysis_v}(b), we superimpose other available limits outlined above. It is observed that our results place a more stringent constraint than most previous limits at low-mass regions. This emphasizes the importance of this study. Specifically, at masses below $m_{Z'}\lesssim 4$ MeV, the results of Run0 and Run1 completely cover the COHERENT limit (combined CsI+LAr analysis), leading to improvements of about one order of magnitude. Also, we obtain more stringent constraints than those derived from nuclear reactor experiments (CONNIE (95\% C.L.), TEXONO ($\overline{\nu}_e e$ scattering), Dresden-II and CONUS+), DD experiment (CDEX-10 (CE$\nu$NS channel))  and other electron-neutrino scattering experiments (CHARM-II ($\overline{\nu}_\mu e$ scattering), LSND ($\nu_e e$ scattering), and NA64). The Run0 limit approximately overlaps the GEMMA and CONUS ($\nu e$ scattering data) limits, while the Run1 limit provides a few times improvement over these.  
The Run0 (Run1) result provides a $25\%$ ($65\%$) improvement over the BOREXINO limit for the region of $m_{Z'}\leq 70$ keV. However, the Run0 and Run1 limits are of the same order of magnitude as the XENONnT limit yet provide no improvement. It is possible that a small region with the allowed bound of $(g-2)_\mu$ survives in the region of $m_{Z'} > 100$ MeV, which is not excluded by our results. 
Additionally, our results are complementary to the BBN, stellar cooling, and collider limits.

We next show our results for the universal light tensor mediator model in Fig. \ref{fig:analysis_t}(a). The upper limit of $g_\mathrm{T}$ reaches $2.8\times 10^{-7}$ for the Run0 data and $1.4\times 10^{-7}$ for the Run1 data in the region of $m_{\mathrm{T}}\leq 1$ keV. The Run1 data leads to an improvement by about 2 times on the limit of $g_\mathrm{T}$ over the Run0 data in the low-mass region. 
We overlaid the existing limits from previous studies in Fig. \ref{fig:analysis_t}(b). Our results significantly cover the limits from COHERENT (combined CsI+LAr analysis) and CDEX-10, improving them by one and two orders of magnitude, respectively. Withal, the obtained results are of the same order of magnitude with the XENONnT limit in the considered parameter space but yet to reach it.
%%%%%%%
\begin{figure*}[htb]
	\centering
	\includegraphics[scale=0.45]{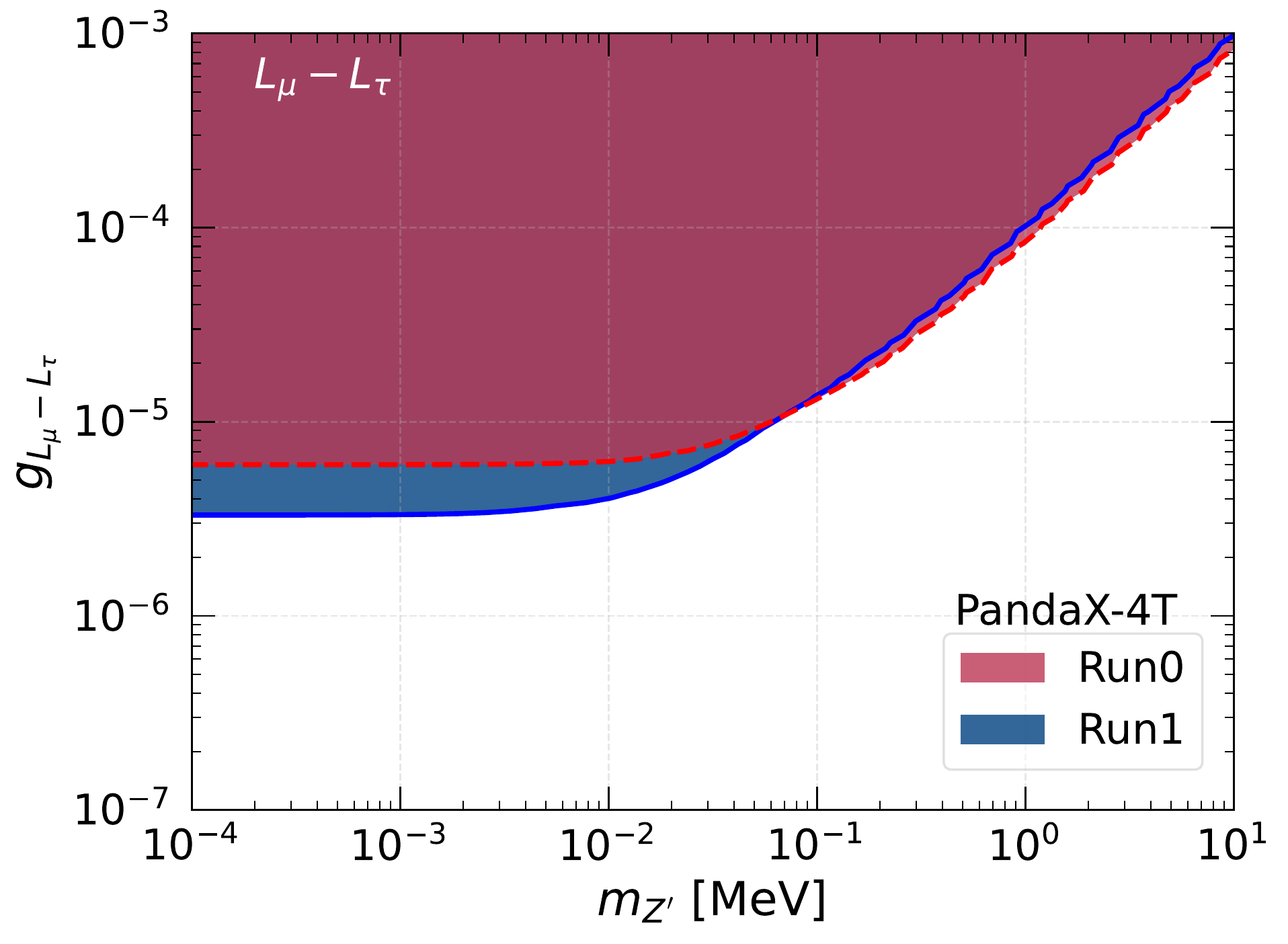}
	\includegraphics[scale=0.45]{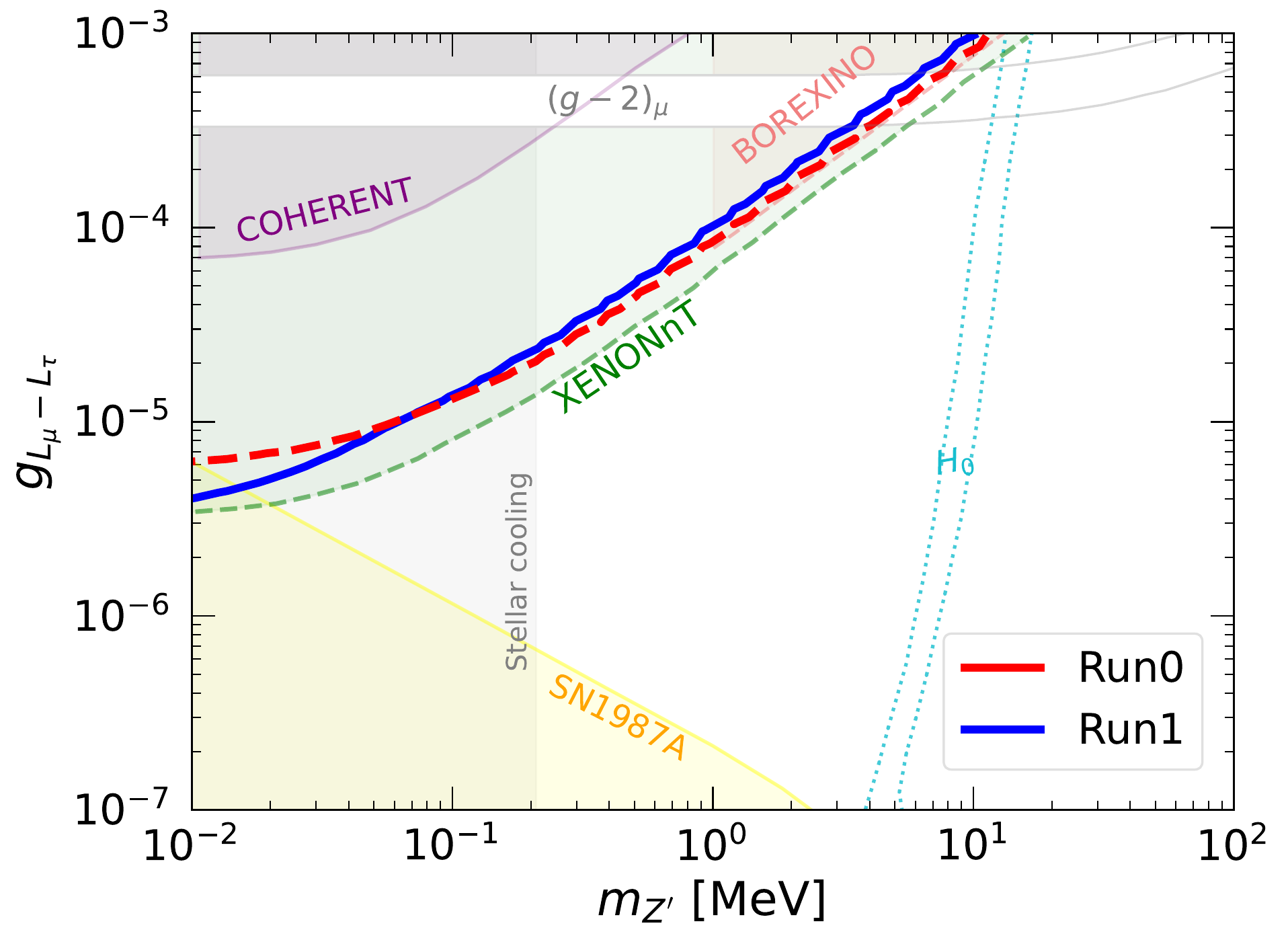}
	\\
    \hspace{10mm} (a) \hspace{83mm} (b)
	\\
	\caption{(a) 90\% C.L. with 2 d.o.f. exclusion regions on the mass-coupling plane of the $L_\mu-L_\tau$ model from the Run0 and Run1 datasets of PandaX-4T. (b) Comparison with existing limits of COHERENT and BOREXINO obtained in Refs. \cite{Melas:2023olz, Gninenko:2020xys}, and XENONnT derived in this work. The limits from supernova (SN1987A) \cite{Croon:2020lrf}, stellar cooling \cite{LiXu:2023} and cosmology (the reported Hubble tension $H_0$) \cite{Escudero:2019gzq} as well as the allowed bound of $(g-2)_\mu$ \cite{Muong-2:2023cdq} are also shown.
	}
	\label{fig:analysis_lmlt}
\end{figure*}
%%%%%%%%
\begin{table*}[ht]
	\caption{90\% C.L. (2 d.o.f.) upper limits on the coupling constants for each light mediator model under consideration. These results are presented for the regions of $m_{\phi,Z',T} < 4$ keV. Here, we present the available limits derived from COHERENT (combined CsI+LAr analysis), BOREXINO and XENONnT. 
	}
	\begin{center}
		\begin{ruledtabular}
		\begin{tabular}{ c c c c c c }
%			\hline
%			\hline
			\multirow{2}{1.5cm}{\textbf{Coupling}} & \multicolumn{2}{c}{\textbf{PandaX-4T (This work)}} &\multirow{2}{4.6cm}{\textbf{COHERENT (CsI+LAr)}} &  \multirow{2}{2cm}{\textbf{BOREXINO}}
			& \multirow{2}{2cm}{\textbf{XENONnT}}
			\\
			\cline{2-3}
			& \textbf{Run0} &\textbf{ Run1} & & &
			\\
			\hline
%			&&&&& \\
			& & & \textit{Universal mediator models}  & &
			\\
			\hline
			$g_\phi$ & $\lesssim 1.1\times 10^{-6}$ & $ \lesssim 8.3\times 10^{-7}$ & $ \lesssim 1.9\times 10^{-5}$ \cite{DeRomeri:2022twg} & 
			$\lesssim1.5\times10^{-6}$ \cite{Coloma:2022umy}& 
			$\lesssim 5.3\times 10^{-7}$ \cite{A:2022acy}
			\\ 
			 $g_{Z'}$ & $\lesssim 5.1\times 10^{-7}$ &  $ \lesssim 2.4\times 10^{-7}$ & $ \lesssim 6.1\times 10^{-6}$ \cite{DeRomeri:2022twg} &
			$ \lesssim 6.8\times 10^{-7}$ \cite{Coloma:2022umy} &
			$ \lesssim 1.9\times 10^{-7}$ \cite{A:2022acy}
			\\ 
			$g_{T}$ & $ \lesssim 2.8\times 10^{-7}$ & $ \lesssim 1.4\times 10^{-7}$ & $ \lesssim 3.5\times 10^{-6}$ \cite{DeRomeri:2022twg} &  
			- &
			$\lesssim 1.0\times 10^{-7}$ \cite{A:2022acy}
		    \\
		    \hline
		    & & & \textit{Lepton flavor-dependent $U(1)'$ models} & & 
		    \\
		    \hline
		    $g_{L_e-L_\mu}$ & $\lesssim 8.2\times 10^{-7}$ &  $ \lesssim 3.9\times 10^{-7}$ & $ \lesssim 6.1\times 10^{-6}$ \cite{Coloma:2022umy} &
		    $ \lesssim 6.2\times 10^{-7}$ \cite{Coloma:2022umy} &
		    $ \lesssim 4.0\times 10^{-7}$~~~
		    \\ 
		    $g_{L_e-L_\tau}$ & $\lesssim 1.2\times 10^{-6}$ &  $ \lesssim 3.7\times 10^{-7}$ & $ \lesssim 7.9\times 10^{-6}$ \cite{Coloma:2022umy} &
		    $ \lesssim 6.4\times 10^{-7}$ \cite{Coloma:2022umy} &
		    $\lesssim 3.8\times 10^{-7}$~~~
		    \\ 
		    $g_{L_\mu-L_\tau}$ & $\lesssim 6.0\times 10^{-6}$ &  $ \lesssim 3.3\times 10^{-6}$ & $ \lesssim 7.0\times 10^{-5}$  \cite{Melas:2023olz}&
		    $ \lesssim 7.0\times 10^{-5}$  \cite{Gninenko:2020xys}&
		    $ \lesssim 3.2\times 10^{-6}$~\multirow{-3.4}{*}{~\rotatebox[origin=c]{-90}{\parbox[r]{1.39cm}{\scriptsize This work}}}
		    \\ 
		\end{tabular}
			\end{ruledtabular}
	\end{center}
	\label{tab:couplings_univ}
\end{table*}
%%
%%%

We finally present the exclusion limits with 90\% C.L. for the $L_e-L_\mu$, $L_e-L_\tau$, and $L_\mu-L_\tau$ models in Figs. \ref{fig:analysis_lelm}(a), \ref{fig:analysis_lelt}(a), and \ref{fig:analysis_lmlt}(a), respectively. The upper limits of $g_{L_e-L_\mu}$ are obtained as $8.2\times 10^{-7}$ and $3.9\times 10^{-7}$ from the Run0 and Run1 datasets, respectively, resulting in around 2.1 times sensitives over each other. Meanwhile, the upper-limits of $g_{L_e-L_\tau}$ reach $1.2\times 10^{-6}$ and $3.8\times 10^{-7}$ for the Run0 and Run1 datasets, respectively. It corresponds to approximately 3.2 times improvement between these two. On the other hand, the upper-limit of $g_{L_\mu-L_\tau}$ from the Run0 data is obtained as $6.0\times 10^{-6}$ and the Run1 data provides $3.3\times 10^{-6}$ corresponding to 1.8 times improvement over the Run0 result. 

In addition, we derive new limits from the current electronic recoil XENONnT \cite{XENON:2022ltv} data for each case, following the $\chi^2$-analysis of Eq.\eqref{eq:chi2}. The $L_e-L_\mu$ and $L_e-L_\tau$ limits are derived for the first time for this data, while the $L_\mu-L_\tau$ limit is compatible with the previous work in Ref. \cite{Melas:2023olz}. The limits of these models have been obtained by a combined analysis of the DD data in Ref. \cite{DeRomeri:2024dbv} but did not provide the separated limit from XENONnT.
We follow the same analysis procedure described in \cite{DeRomeri:2024dbv} for dealing with XENONnT data.

In Figs. \ref{fig:analysis_lelm}(b), \ref{fig:analysis_lelt}(b), and \ref{fig:analysis_lmlt}(b), comparisons of our results with previous limits are shown in the region of $10 \text{ keV}\leq m_{Z'}\leq 100 \text{ MeV}$ for the $L_e-L_\mu$, $L_e-L_\tau$, and $L_\mu-L_\tau$ models, respectively. We note that for XENONnT we provide our derived limits for each case.
For the $L_e-L_\mu$ and $L_e-L_\tau$ models, the previous limits are also presented from solar neutrino facility (BOREXINO), nuclear reactors (TEXONO, CONUS, Dresden-II), stopped-pion source (COHERENT-combined CsI+LAr analysis), BBN, stellar cooling and oscillation data. 
For the $L_\mu-L_\tau$ model, additional limits from supernova (SN1987A) and cosmology (H0) are presented.

We now discuss the comparisons for the $L_e-L_\mu$ and $L_e-L_\tau$ models. Our results provide up to one order of magnitude and three times improvement over COHERENT and Dresden-II limits, respectively. 
The TEXONO limit is covered for $m_{Z'}\leq 0.2$ MeV by providing an improvement of up to 6 times. Compared to the CONUS limit, our results are competitive in the considered mass region, passing the limit as $m_{Z'}\leq 20$ keV. Furthermore, the Run1 data is more stringent in the region of $m_{Z'}\leq 20$ keV for the $L_e-L_\mu$ model and $m_{Z'}\leq 30$ keV for the $L_e-L_\tau$ model. Regarding the XENONnT limits, the Run1 limits surpass them as $m_{Z'}\lesssim 20$ keV for both models. Meanwhile, our results are yet to reach oscillation limits, but provide complementarity to the BBN and stellar cooling bounds. 

Regarding the $L_\mu-L_\tau$ models, our results provide a more stringent limit than COHERENT and BOREXINO, albeit a few times less stringent than XENONnT in the considered parameter space. Meanwhile, the limits of SN1987A and stellar cooling are reached in the low-mass region. 
Although its low-mass area is mostly excluded, some part of the allowed band of $(g-2)_\mu$ survives in the region of $m_{Z'}>4$ MeV. We also show the region that explains the Hubble tension $H_0$ for $\Delta N_{\text{eff}}\sim 0.2 - 0.5$ taken from Ref. \cite{Escudero:2019gzq}.

In Table \ref{tab:couplings_univ}, we list our results for completeness. These limits can be read directly from Figs. \ref{fig:analysis_s}, \ref{fig:analysis_v}, and \ref{fig:analysis_t} for universal mediator models as well as from Figs. \ref{fig:analysis_lelm}, \ref{fig:analysis_lelt}, and \ref{fig:analysis_lmlt} for the lepton flavor-dependent of $L_e-L_\mu$, $L_e-L_\tau$, and $L_\mu-L_\tau$ models, respectively. For comparison, we also include previous limits derived from COHERENT, BOREXINO, and XENONnT data. Our limits are more sensitive than COHERENT (stopped-pion source) and BOREXINO (solar neutrino) limits and competitive with the XENONnT (direct detection) limit.

As a general remark, it should be emphasized that in all cases PandaX-4T is expected to provide improved constraints to those obtained from low-energy solar, stopped-pion, and reactor experiments and be complementary to high-energy collider and astrophysical searches. The PandaX-4T, as a direct detection experiment, can provide more sensitive constraints than many previously derived results thanks to its low recoil energy threshold, experimental exposure, and target mass. We can notice this, particularly for COHERENT which has a recoil energy threshold at around $\sim5$ keV. The PandaX-4T datasets provide up to one order of magnitude improvement over the COHERENT limit. Furthermore, the limits obtained from nuclear reactor experiments are better than those from stopped-pion sources, while competing with the PandaX-4T limit. For example, TEXONO, CONNIE, CONUS+ and Dresden-II have higher sensitivities at larger masses, due to the larger measured recoil energy, while the CONUS with sub-keV threshold can provide a competitive background at low-mass regions. Compared to the solar neutrino experiment of BOREXINO, the PandaX-4T datasets provide more stringent limits in the low-mass region, whilst it dominates in the large-mass region due to the larger recoil energy measured in the detector. The PandaX-4T limits are able to reach other DD results for some models at low mass scale, while yet to reach for others.

\section{Summary and Conclusions}\label{sec:summ}
We have investigated new physics effects from the universal light mediator and lepton flavor-dependent $U(1)'$ models through neutrino-electron scatterings with solar neutrinos using the recent Run0 and Run1 datasets of the PandaX-4T experiment. The universal mediator models consist of scalar, vector, and tensor interactions allowed by Lorentz invariance and involving light mediators. A vector mediator from the $U(1)_{L_e-L_{\mu}}, U(1)_{L_e-L_{\tau}},$ and $U(1)_{L_\mu-L_{\tau}}$ models has been considered, where it differs from the universal vector case with their vector charge. All of these light mediators can play a role as possible new physics signatures as neutrinos interact with matter, namely electron components of targets in DM DD facilities.

We have calculated the event-rate contribution of each model to neutrino-electron scattering using solar neutrino fluxes. In general, these models may yield deviations from the SM prediction. The new physics effects from the light mediators have emerged in the low-scale electron recoil energy, indicating the need for enhancing detector sensitivity in this region. We have derived new and updated constraints on the coupling-mass plane of the considered models from the recent PandaX-4T Run0 and Run1 datasets. It is observed that the Run1 data is a few times more sensitive than the Run0 data. We have also compared our results with the available limits in the literature, derived from various experimental probes. Our results show an improvement over most of the existing limits for the considered models.
The PandaX-4T, being a direct detection experiment, provides more stringent limits than the stopped pion neutrino sources and nuclear reactor experiments due to its lower energy threshold. Compared to other direct detection experiments, the PandaX-4T results are competitive or even better than others despite its relatively small exposure. Concerning the solar neutrino result, the PandaX-4T (Run1) data puts more competitive and improved limits at low-mass regions. These results also complement the available constraints from various collider experiments, supernova (SN1987A), stellar cooling, and BBN, as well as the allowed band of $(g-2)_\mu$.

All in all, we have demonstrated that neutrino-electron scattering with solar neutrinos in a direct detection facility can provide more competitive and improved constraints according to existing results.  We anticipate that our results could be used to complement activities of hunting new physics signatures with light mediators from various models with recent and future experimental advancements related to solar neutrino and DM DD experiments.

\section{Acknowledgments} 
This work was supported by the Scientific and Technological Research Council of Türkiye (TUBITAK) under the project no: 124F416.


\begin{thebibliography}{99}
%
\bibitem{tHooft:1971ucy}
G.~'t Hooft,
%``Predictions for neutrino - electron cross-sections in Weinberg's model of weak interactions,''
\href{https://doi.org/10.1016/0370-2693(71)90050-5}{Phys. Lett. B \textbf{37}, 195-196 (1971)}.

\bibitem{Bahcall:1995mm}
J.~N.~Bahcall, M.~Kamionkowski and A.~Sirlin,
%``Solar neutrinos: Radiative corrections in neutrino - electron scattering experiments,''
\href{https://doi.org/10.1103/PhysRevD.51.6146}{Phys. Rev. D \textbf{51}, 6146-6158 (1995)}.
%[arXiv:astro-ph/9502003 [astro-ph]].

%\cite{Davis:1968cp}
\bibitem{Davis:1968cp}
R.~Davis, Jr., D.~S.~Harmer and K.~C.~Hoffman,
%``Search for neutrinos from the sun,''
\href{http://10.1103/PhysRevLett.20.1205}{{Phys. Rev. Lett.} \textbf{20}, 1205-1209 (1968)}.


%\cite{GALLEX:1992gcp}
\bibitem{GALLEX:1992gcp}
P.~Anselmann \textit{et al.} (GALLEX Collaboration),
%``Solar neutrinos observed by GALLEX at Gran Sasso.,''
\href{https://doi.org/10.1016/0370-2693(92)91521-A}{{Phys. Lett. B} \textbf{285}, 376-389 (1992)}.
%680 citations counted in INSPIRE as of 20 Nov 2022

%\cite{Cleveland:1998nv}
\bibitem{Cleveland:1998nv}
B.~T.~Cleveland, T.~Daily, R.~Davis Jr., J.~R.~Distel, K.~Lande, C.~K.~Lee, P.~S.~Wildenhain and J.~Ullman,
%``Measurement of the solar electron neutrino flux with the Homestake chlorine detector,''
\href{http://doi.org/10.1086/305343}{{Astrophys. J.} \textbf{496}, 505-526 (1998)}.
%2867 citations counted in INSPIRE as of 20 Nov 2022

%\cite{SAGE:1999uje}
\bibitem{SAGE:1999uje}
J.~N.~Abdurashitov \textit{et al.} (SAGE Collaboration),
%``Measurement of the solar neutrino capture rate by SAGE and implications for neutrino oscillations in vacuum,''
\href{https://doi.org/10.1103/PhysRevLett.83.4686}{{Phys. Rev. Lett.} \textbf{83}, 4686-4689 (1999)}. % [arXiv:astro-ph/9907131 [astro-ph]].
%251 citations counted in INSPIRE as of 20 Nov 2022

%\cite{SNO:2002tuh}
\bibitem{SNO:2002tuh}
Q.~R.~Ahmad \textit{et al.} (SNO Collaboration),
%``Direct evidence for neutrino flavor transformation from neutral current interactions in the Sudbury Neutrino Observatory,''
\href{https://doi.org/10.1103/PhysRevLett.89.011301}{{Phys. Rev. Lett.} \textbf{89}, 011301 (2002)}.
%[arXiv:nucl-ex/0204008 [nucl-ex]].
%4217 citations counted in INSPIRE as of 20 Nov 2022

%\cite{SNO:2003bmh}
\bibitem{SNO:2003bmh}
S.~N.~Ahmed \textit{et al.} (SNO Collaboration),
%``Measurement of the total active B-8 solar neutrino flux at the Sudbury Neutrino Observatory with enhanced neutral current sensitivity,''
\href{https://doi.org/10.1103/PhysRevLett.92.181301}{{Phys. Rev. Lett.} \textbf{92}, 181301  (2004)}.
%[arXiv:nucl-ex/0309004 [nucl-ex]].
%1245 citations counted in INSPIRE as of 20 Nov 2022

%\cite{SNO:2008gqy}
\bibitem{SNO:2008gqy}
B.~Aharmim \textit{et al.} (SNO Collaboration),
%``An Independent Measurement of the Total Active B-8 Solar Neutrino Flux Using an Array of He-3 Proportional Counters at the Sudbury Neutrino Observatory,''
\href{https://doi.org/10.1103/PhysRevLett.101.111301}{{Phys. Rev. Lett.} \textbf{101}, 111301  (2008)}.
%[arXiv:0806.0989 [nucl-ex]].
%433 citations counted in INSPIRE as of 20 Nov 2022

%\cite{Kamiokande-II:1989hkh}
\bibitem{Kamiokande-II:1989hkh}
K.~S.~Hirata \textit{et al.} (Kamiokande-II Collaboration),
%``Observation of B-8 Solar Neutrinos in the Kamiokande-II Detector,''
\href{https://doi.org/10.1103/PhysRevLett.63.16}{{Phys. Rev. Lett.} \textbf{63}, 16 (1989)}.
%638 citations counted in INSPIRE as of 20 Nov 2022

%\cite{Super-Kamiokande:2001ljr}
\bibitem{Super-Kamiokande:2001ljr}
S.~Fukuda \textit{et al.} (Super-Kamiokande Collaboration),
%``Solar B-8 and hep neutrino measurements from 1258 days of Super-Kamiokande data,''
\href{https://doi.org/10.1103/PhysRevLett.86.5651}{{Phys. Rev. Lett.} \textbf{86}, 5651-5655 (2001)}.
%[arXiv:hep-ex/0103032 [hep-ex]].
%1443 citations counted in INSPIRE as of 20 Nov 2022

%\cite{Borexino:2007kvk}
\bibitem{Borexino:2007kvk}
C.~Arpesella \textit{et al.} (Borexino Collaboration),
%``First real time detection of Be-7 solar neutrinos by Borexino,''
\href{https://doi.org/10.1016/j.physletb.2007.09.054}{{Phys. Lett. B} \textbf{658}, 101-108 (2008)}.
%[arXiv:0708.2251 [astro-ph]].
%326 citations counted in INSPIRE as of 20 Nov 2022

%\cite{Borexino:2008fkj}
\bibitem{Borexino:2008fkj}
G.~Bellini \textit{et al.} (Borexino Collaboration),
%``Measurement of the solar 8B neutrino rate with a liquid scintillator target and 3 MeV energy threshold in the Borexino detector,''
\href{https://doi.org/10.1103/PhysRevD.82.033006}{{Phys. Rev. D} \textbf{82}, 033006 (2010)}.
%[arXiv:0808.2868 [astro-ph]].
%415 citations counted in INSPIRE as of 20 Nov 2022

%\cite{Bahcall:2000nu}
\bibitem{Bahcall:2000nu}
J.~N.~Bahcall, M.~H.~Pinsonneault and S.~Basu,
%``Solar models: Current epoch and time dependences, neutrinos, and helioseismological properties,''
\href{https://doi.org/10.1086/321493}{{Astrophys. J.} \textbf{555}, 990-1012 (2001)}.
%[arXiv:astro-ph/0010346 [astro-ph]].
%1004 citations counted in INSPIRE as of 20 Nov 2022

%\cite{Serenelli:2016dgz}
\bibitem{Serenelli:2016dgz}
A.~Serenelli,
%``Alive and well: a short review about standard solar models,''
\href{https://doi.org/10.1140/epja/i2016-16078-1}{{Eur. Phys. J. A} \textbf{52}, 78 (2016)}.
%[arXiv:1601.07179 [astro-ph.SR]].
%51 citations counted in INSPIRE as of 20 Nov 2022

%\cite{Vinyoles:2016djt}
\bibitem{Vinyoles:2016djt}
N.~Vinyoles, A.~M.~Serenelli, F.~L.~Villante, S.~Basu, J.~Bergstr\"om, M.~C.~Gonzalez-Garcia, M.~Maltoni, C.~Pe\~na-Garay and N.~Song,
%``A new Generation of Standard Solar Models,''
\href{https://doi.org/10.3847/1538-4357/835/2/202}{{Astrophys. J.} \textbf{835}, 202 (2017)}.
%[arXiv:1611.09867 [astro-ph.SR]].
%230 citations counted in INSPIRE as of 20 Nov 2022

\bibitem{Vitagliano:2019yzm}
E.~Vitagliano, I.~Tamborra and G.~Raffelt,
%``Grand Unified Neutrino Spectrum at Earth: Sources and Spectral Components,''
\href{http://doi.org/10.1103/RevModPhys.92.045006}{{Rev. Mod. Phys.} \textbf{92}, 45006 (2020)}.
%[arXiv:1910.11878 [astro-ph.HE]].

%\cite{Bahcall:2004pz}
\bibitem{Bahcall:2004pz}
J.~N.~Bahcall, A.~M.~Serenelli and S.~Basu,
%``New solar opacities, abundances, helioseismology, and neutrino fluxes,''
\href{http://doi.org/10.1086/428929}{{Astrophys. J. Lett.} \textbf{621}, L85-L88 (2005)}.
%[arXiv:astro-ph/0412440 [astro-ph]].
%750 citations counted in INSPIRE as of 01 Apr 2023

%\cite{Bahcall:2004mq}
\bibitem{Bahcall:2004mq}
J.~N.~Bahcall and A.~M.~Serenelli,
%``How do uncertainties in the surface chemical abundances of the Sun affect the predicted solar neutrino fluxes?,''
\href{http://doi.org/10.1086/429883}{{Astrophys. J} \textbf{626}, 530 (2005)}.
%[arXiv:astro-ph/0412096 [astro-ph]].
%81 citations counted in INSPIRE as of 01 Apr 2023

%\cite{Cerdeno:2016sfi}
\bibitem{Cerdeno:2016sfi}
D.~G.~Cerde\~no, M.~Fairbairn, T.~Jubb, P.~A.~N.~Machado, A.~C.~Vincent and C.~B\oe{}hm,
%``Physics from solar neutrinos in dark matter direct detection experiments,'' 
\href{https://doi.org/10.1007/JHEP05(2016)118}{ J. High Energ. Phys. \textbf{2016}, 118 (2016)} [erratum: \href{https://doi.org/10.1007/JHEP09(2016)048}{ J. High Energ. Phys. \textbf{2016}, 48 (2016)}]. %[arXiv:1604.01025 [hep-ph]].
%108 citations counted in INSPIRE as of 20 Nov 2022

%\cite{Drukier:1984vhf}
\bibitem{Drukier:1984vhf}
A.~Drukier and L.~Stodolsky,
%``Principles and Applications of a Neutral Current Detector for Neutrino Physics and Astronomy,''
\href{https://doi.org/10.1103/PhysRevD.30.2295}{{Phys. Rev. D } \textbf{30}, 2295 (1984)}.
%487 citations counted in INSPIRE as of 20 Nov 2022

%\cite{Goodman:1984dc}
\bibitem{Goodman:1984dc}
M.~W.~Goodman and E.~Witten,
%``Detectability of Certain Dark Matter Candidates,''
\href{https://doi.org/10.1103/PhysRevD.31.3059}{{Phys. Rev. D } \textbf{31}, 3059 (1985)}.
%1426 citations counted in INSPIRE as of 20 Nov 2022


%\cite{Drukier:1986tm}
\bibitem{Drukier:1986tm}
A.~K.~Drukier, K.~Freese and D.~N.~Spergel,
%``Detecting Cold Dark Matter Candidates,''
\href{https://doi.org/10.1103/PhysRevD.33.3495}{{Phys. Rev. D } \textbf{33}, 3495-3508 (1986)}.
%977 citations counted in INSPIRE as of 20 Nov 2022



%\cite{CDEX:2022mlp}
\bibitem{CDEX:2022mlp}
X.~P.~Geng \textit{et al.} (CDEX Collaboration),
%``Search for exotic neutrino interactions using solar neutrinos in the CDEX-10 experiment,''
\href{https://doi.org/10.1103/PhysRevD.107.112002}{{Phys. Rev. D}\textbf{107}, 112002 (2023)}.
%[arXiv:2210.01604 [hep-ex]].
%0 citations counted in INSPIRE as of 07 Nov 2022


%\cite{XENON:2020kmp}
\bibitem{XENON:2020kmp}
E.~Aprile \textit{et al.} (XENON Collaboration),
%``Projected WIMP sensitivity of the XENONnT dark matter experiment,''
\href{http://doi.org/10.1088/1475-7516/2020/11/031}{{JCAP} \textbf{11}, 031 (2020)}. %[arXiv:2007.08796 [physics.ins-det]].
%233 citations counted in INSPIRE as of 20 Nov 2022

\bibitem{XENON:2022ltv}
E.~Aprile \textit{et al.} (XENON Collaboration),
%``Search for New Physics in Electronic Recoil Data from XENONnT,''
\href{http://doi.org/10.1103/PhysRevLett.129.161805}{Phys. Rev. Lett. \textbf{129}, 161805 (2022)}.
%[arXiv:2207.11330 [hep-ex]].

%\cite{LUX:2015abn}
\bibitem{LUX:2015abn}
D.~S.~Akerib \textit{et al.} (LUX-ZEPLIN Collaboration),
%``Improved Limits on Scattering of Weakly Interacting Massive Particles from Reanalysis of 2013 LUX Data,''
\href{http://doi.org/10.1103/PhysRevLett.116.161301}{{Phys. Rev. Lett.} \textbf{116}, 161301  (2016)}.% [arXiv:1512.03506 [astro-ph.CO]].
%549 citations counted in INSPIRE as of 20 Nov 2022

%\cite{LZ:2018qzl}
\bibitem{LZ:2018qzl}
D.~S.~Akerib \textit{et al.} (LUX-ZEPLIN Collaboration),
%``Projected WIMP sensitivity of the LUX-ZEPLIN dark matter experiment,''
\href{http://doi.org/10.1103/PhysRevD.101.052002}{{Phys. Rev. D} \textbf{101}, 052002 (2020)}. % [arXiv:1802.06039 [astro-ph.IM]].
%348 citations counted in INSPIRE as of 20 Nov 2022

\bibitem{LZ:2023ja}
J. Aalbers et al. (LUX-ZEPLIN Collaboration),
%"Search for new physics in low-energy electron recoils from the first LZ exposure,"
\href{https://doi.org/10.1103/PhysRevD.108.072006?}{{Phys. Rev. D} \textbf{108}, 072006 (2023)}.

%\cite{DarkSide:2018kuk}
\bibitem{DarkSide:2018kuk}
P.~Agnes \textit{et al.} (DarkSide Collaboration),
%``DarkSide-50 532-day Dark Matter Search with Low-Radioactivity Argon,''
\href{http://doi.org/10.1103/PhysRevD.98.102006}{{Phys. Rev. D} \textbf{98}, 102006 (2018)}. %[arXiv:1802.07198 [astro-ph.CO]].
%215 citations counted in INSPIRE as of 20 Nov 2022

\bibitem{PandaX:2014mem}
X.~Cao \textit{et al.} (PandaX Collaboration),
%``PandaX: A Liquid Xenon Dark Matter Experiment at CJPL,''
\href{http://doi.org/10.1007/s11433-014-5521-2}{Sci. China Phys. Mech. Astron. \textbf{57}, 1476-1494 (2014)}.
%[arXiv:1405.2882 [physics.ins-det]].

%\cite{PandaX-II:2017hlx}
\bibitem{PandaX-II:2017hlx}
X.~Cui \textit{et al.} (PandaX-II Collaboration),
%``Dark Matter Results From 54-Ton-Day Exposure of PandaX-II Experiment,''
\href{http://doi.org/10.1103/PhysRevLett.119.181302}{{Phys. Rev. Lett.} \textbf{119}, 181302 (2017)}. % [arXiv:1708.06917 [astro-ph.CO]].
%981 citations counted in INSPIRE as of 20 Nov 2022

\bibitem{DARWIN:2020bnc}
J.~Aalbers \textit{et al.} (DARWIN Collaboration),
%``Solar neutrino detection sensitivity in DARWIN via electron scattering,''
\href{http://doi.org/10.1140/epjc/s10052-020-08602-7}{Eur. Phys. J. C \textbf{80}, 1133  (2020)}.
%[arXiv:2006.03114 [physics.ins-det]].

\bibitem{DARKSIDE20K:2021}
V.~Pesudo \textit{et al.} (DarkSide-20k Collaboration),
%``Measurement of the underground argon radiopurity for Dark Matter direct searches,''
\href{http://doi.org/10.1088/1742-6596/2156/1/012043}{ J. Phys.: Conf. Ser. \textbf{2156}, 012043  (2021)}.


\bibitem{PandaXT:20248b}
Z.~Bo \textit{et al.} (PandaX Collaboration),
%``First Indication of Solar $^{8}\mathrm{B}$ Neutrinos through Coherent Elastic Neutrino-Nucleus Scattering in PandaX-4T,''
\href{https://link.aps.org/doi/10.1103/PhysRevLett.133.191002}{Phys. Rev. Lett. \textbf{133}, 191001 (2024)}.
%[arXiv:2408.02877 [nucl-ex]].

\bibitem{XENONnT:20248b}
E.~Aprile \textit{et al.} (XENON Collaboration),
%``First Indication of Solar $^{8}\mathrm{B}$ Neutrinos via Coherent Elastic Neutrino-Nucleus Scattering with XENONnT,''
\href{https://link.aps.org/doi/10.1103/PhysRevLett.133.191002}{Phys. Rev. Lett. \textbf{133}, 191002 (2024)}.
%[arXiv:2408.02877 [nucl-ex]].

\bibitem{PandaX:2022ood}
D.~Zhang \textit{et al.} (PandaX Collaboration),
%``Search for Light Fermionic Dark Matter Absorption on Electrons in PandaX-4T,''
\href{http://doi.org/10.1103/PhysRevLett.129.161804}{Phys. Rev. Lett. \textbf{129}, 161804 (2022)}.
%[arXiv:2206.02339 [hep-ex]].

\bibitem{PandaX:2018wtu}
H.~Zhang \textit{et al.} (PandaX Collaboration),
%``Dark matter direct search sensitivity of the PandaX-4T experiment,''
\href{http://dx.doi.org/10.1007/s11433-018-9259-0}{Sci. China Phys. Mech. Astron. \textbf{62}, 31011 (2019)}.
%[arXiv:1806.02229 [physics.ins-det]].

\bibitem{PandaX:2024med}
Y.~Luo \textit{et al.} (PandaX Collaboration),
%``Signal response model in PandaX-4T,''
\href{http://dx.doi.org/10.1103/PhysRevD.110.023029}{Phys. Rev. D \textbf{110}, 023029 (2024)}.
%[arXiv:2403.04239 [physics.ins-det]].

\bibitem{PandaX:2024zbo}
Z.~Bo \textit{et al.} (PandaX Collaboration),
%``Dark Matter Search Results from 1.54 Tonne⋅Year Exposure of PandaX-4T,''
\href{https://doi.org/10.1103/PhysRevLett.134.011805}{Phys. Rev. Lett. \textbf{134}, 011805 (2025)}.
%\href{https://doi.org/10.48550/arXiv.2408.00664}{arXiv:2408.00664 [hep-ex]}.

\bibitem{PandaX:2024cic}
X.~Zeng \textit{et al.} (PandaX Collaboration),
%``Exploring New Physics with PandaX-4T Low Energy Electronic Recoil Data,''
\href{https://doi.org/10.1103/PhysRevLett.134.041001}{Phys. Rev. Lett. \textbf{134}, 041001 (2025)}.
%\href{https://arxiv.org/pdf/2408.07641}{arXiv:2408.07641 [hep-ex]}.


\bibitem{Arkani-Hamed:2008hhe}
N.~Arkani-Hamed, D.~P.~Finkbeiner, T.~R.~Slatyer and N.~Weiner,
%``A Theory of Dark Matter,''
\href{https://doi.org/10.1103/PhysRevD.79.015014}{Phys. Rev. D \textbf{79}, 015014 (2009)}.
%[arXiv:0810.0713 [hep-ph]].

\bibitem{Dasgupta:2021ies}
B.~Dasgupta and J.~Kopp,
%``Sterile Neutrinos,''
\href{https://doi.org/10.1016/j.physrep.2021.06.002}{Phys. Rept. \textbf{928}, 1-63 (2021)}.
%[arXiv:2106.05913 [hep-ph]].

\bibitem{Brdar:2018}
V. Brdar,  W. Rodejohann and X. J. Xu, %"Producing a new fermion in coherent elastic neutrino-nucleus scattering: from neutrino mass to dark matter." 
\href{https://doi.org/10.1007/JHEP12(2018)024}{J. High Energ. Phys. \textbf{2018}, 24 (2018)}


\bibitem{Essig:2013lka} R. Essig \textit{et al.}, %Working Group Report: New Light Weakly Coupled Particles, Report No. YITP-SB-36,
\href{https://doi.org/10.48550/arXiv.1311.0029}{arXiv:1311.0029 [hep-ph]}.


%\cite{Abdallah:2015ter}
\bibitem{Abdallah:2015ter}
J.~Abdallah, H.~Araujo, A.~Arbey, A.~Ashkenazi, A.~Belyaev, J.~Berger, C.~Boehm, A.~Boveia, A.~Brennan and J.~Brooke,
%``Simplified Models for Dark Matter Searches at the LHC,''
\href{https://doi.org/10.1016/j.dark.2015.08.001}{{Phys. Dark Univ.} \textbf{9-10}, 8-23 (2015)}. %[arXiv:1506.03116 [hep-ph]].
%381 citations counted in INSPIRE as of 20 Nov 2022
 

\bibitem{Mohapatra:1980qe}
R.~N.~Mohapatra and R.~E.~Marshak,
%``Local B-L Symmetry of Electroweak Interactions, Majorana Neutrinos and Neutron Oscillations,''
\href{https://doi.org/10.1103/PhysRevLett.44.1316}{Phys. Rev. Lett. \textbf{44}, 1316-1319  (1980)	[erratum: Phys. Rev. Lett. \textbf{44}, 1643 (1980)]}. 

%\cite{He:1991qd}
\bibitem{He:1991qd}
X.~G.~He, G.~C.~Joshi, H.~Lew and R.~R.~Volkas,
%``Simplest Z-prime model,''
\href{https://doi.org/10.1103/PhysRevD.44.2118}{Phys. Rev. D \textbf{44}, 2118-2132 (1991)}.
%426 citations counted in INSPIRE as of 28 Sep 2024

%\cite{Melas:2023olz} DUNE
\bibitem{Melas:2023olz}
P.~Melas, D.~K.~Papoulias and N.~Saoulidou,
%``Probing generalized neutrino interactions with the DUNE Near Detector,''
\href{http://doi.org/10.1007/JHEP07(2023)190}{ J. High Energ. Phys. \textbf{2023}, 190 (2023)}.
%[arXiv:2303.07094 [hep-ph]].
%3 citations counted in INSPIRE as of 15 Sep 2023

\bibitem{CONUS:2021dwh}
H.~Bonet \textit{et al.} (CONUS Collaboration),
%``Novel constraints on neutrino physics beyond the standard model from the CONUS experiment,''
\href{https://doi.org/10.1007/JHEP05(2022)085}{ J. High Energ. Phys. \textbf{2022}, 85 (2022)}. 
%[arXiv:2110.02174 [hep-ph]].

\bibitem{Lindner:2024eng}
M.~Lindner, T.~Rink and M.~Sen,
%``Light vector bosons and the weak mixing angle in the light of future germanium-based reactor CE\ensuremath{\nu}NS experiments,''
\href{https://doi.org/10.1007/JHEP08(2024)171}{J. High Energ. Phys. \textbf{2024}, 171  (2024)}. 
%[arXiv:2401.13025 [hep-ph]].

\bibitem{Chattaraj:2025}
 A. Chattaraj, A. Majumdar and R. Srivastava, Probing Standard Model and Beyond with Reactor CE$\nu$NS Data of CONUS+ experiment, 
 \href{https://doi.org/10.48550/arXiv.2501.12441}{arXiv:2501.12441 [hep-ph]}.

%\cite{Farzan:2018gtr}
\bibitem{Farzan:2018gtr}
Y.~Farzan M.~Lindner, W.~Rodejohann and X.~J.~Xu,
%``Probing neutrino coupling to a light scalar with coherent neutrino scattering,''
\href{https://doi.org/10.1007/JHEP05(2018)066}{ J. High Energ. Phys.  \textbf{2018}, 66 (2018)}.
%[arXiv:1802.05171 [hep-ph]].
%105 citations counted in INSPIRE as of 20 Nov 2022

\bibitem{DeRomeri:2022twg}
V.~De Romeri, O.~G.~Miranda, D.~K.~Papoulias, G.~Sanchez Garcia, M.~T\'ortola and J.~W.~F.~Valle,
%``Physics implications of a combined analysis of COHERENT CsI and LAr data,''
\href{https://doi.org/10.1007/JHEP04(2023)035}{J. High Energ. Phys. \textbf{2023}, 35 (2023)}. 
%[arXiv:2211.11905 [hep-ph]].

%\cite{Cadeddu:2020nbr}
\bibitem{Cadeddu:2020nbr}
M.~Cadeddu, N.~Cargioli, F.~Dordei, C.~Giunti, Y.~F.~Li, E.~Picciau and Y.~Y.~Zhang,
%``Constraints on light vector mediators through coherent elastic neutrino nucleus scattering data from COHERENT,''
\href{http://doi.org/10.1007/JHEP01(2021)116}{ J. High Energ. Phys. \textbf{2021}, 116 (2021)}.
%[arXiv:2008.05022 [hep-ph]].


%\cite{Demirci:2021zci}
\bibitem{Demirci:2021zci}
M.~Demirci and M.~F.~Mustamin, 
%``Probing Light New Mediators on Coherent Elastic Neutrino-Nucleus Scattering,''
\href{https://doi.org/10.31526/ACP.BSM-2021.31}{{Andromeda Proceedings}, BSM21 (2021)}.
%doi:10.31526/ACP.BSM-2021.31
%2 citations counted in INSPIRE as of 17 Sep 2023

%\cite{AtzoriCorona:2022moj}
\bibitem{AtzoriCorona:2022moj}
M.~Atzori Corona, M.~Cadeddu, N.~Cargioli, F.~Dordei, C.~Giunti, Y.~F.~Li, E.~Picciau, C.~A.~Ternes and Y.~Y.~Zhang,
%``Probing light mediators and (g \ensuremath{-} 2)$_{μ}$ through detection of coherent elastic neutrino nucleus scattering at COHERENT,''
\href{http://doi.org/10.1007/JHEP05(2022)109}{J. High Energ. Phys. \textbf{2022}, 109 (2022)}.
%[arXiv:2202.11002 [hep-ph]].
%31 citations counted in INSPIRE as of 06 Sep 2023


\bibitem{Coloma:2022avw}
P.~Coloma, I.~Esteban, M.~C.~Gonzalez-Garcia, L.~Larizgoitia, F.~Monrabal and S.~Palomares-Ruiz,
%``Bounds on new physics with data of the Dresden-II reactor experiment and COHERENT,''
\href{http://doi.org/10.1007/JHEP05(2022)037}{J. High Energ. Phys. \textbf{2022}, 37 (2022)}.
%[arXiv:2202.10829 [hep-ph]].

\bibitem{Coloma:2020gfv}
P.~Coloma, M.~C.~Gonzalez-Garcia and M.~Maltoni,
%``Neutrino oscillation constraints on U(1)' models: from non-standard interactions to long-range forces,''
\href{https://doi.org/10.1007/JHEP01(2021)114}{ J. High Energ. Phys. \textbf{2021}, 114 (2021)}	\href{https://doi.org/10.1007/JHEP11(2022)115}{[erratum: J. High Energ. Phys. \textbf{2022}, 115 (2022)]}.
%[arXiv:2009.14220 [hep-ph]].

%\cite{Coloma:2022umy}
\bibitem{Coloma:2022umy}
P.~Coloma, M.~C.~Gonzalez-Garcia, M.~Maltoni, J.~P.~Pinheiro and S.~Urrea,
%``Constraining new physics with Borexino Phase-II spectral data,''
\href{http://doi.org/10.1007/JHEP07(2022)138}{J. High Energ. Phys. \textbf{2022}, 138 (2022)}. %[arXiv:2204.03011 [hep-ph]].
%6 citations counted in INSPIRE as of 07 Nov 2022


\bibitem{Gninenko:2020xys}
S.~Gninenko and D.~Gorbunov,
%``Refining constraints from Borexino measurements on a light Z'-boson coupled to L\ensuremath{\mu}-L\ensuremath{\tau} current,''
\href{http://doi.org/10.1016/j.physletb.2021.136739}{Phys. Lett. B \textbf{823}, 136739 (2021)}.
%[arXiv:2007.16098 [hep-ph]].

%\cite{Boehm:2020ltd}
\bibitem{Boehm:2020ltd}
C.~Boehm, D.~G.~Cerde\~no, M.~Fairbairn, P.~A.~N.~Machado and A.~C.~Vincent,
%``Light new physics in XENON1T,''
\href{http://10.1103/PhysRevD.102.115013}{{Phys. Rev. D} \textbf{102}, 115013 (2020)}.
%[arXiv:2006.11250 [hep-ph]].
%92 citations counted in INSPIRE as of 20 Nov 2022


%\cite{Schwemberger:2022fjl}
\bibitem{Schwemberger:2022fjl}
T.~Schwemberger and T.~T.~Yu,
%``Detecting beyond the standard model interactions of solar neutrinos in low-threshold dark matter detectors,''
\href{https://doi.org/10.1103/PhysRevD.106.015002}{{Phys. Rev. D} \textbf{106}, 015002 (2022)}. % [arXiv:2202.01254 [hep-ph]].
%[\href{https://arXiv.org/abs/2202.01254}{arXiv:2202.01254 [hep-ph]}.
%7 citations counted in INSPIRE as of 20 Nov 2022

%\cite{A:2022acy}
\bibitem{A:2022acy}
S.~K.~A., A.~Majumdar, D.~K.~Papoulias, H.~Prajapati and R.~Srivastava,
%``Implications of first LZ and XENONnT results: A comparative study of neutrino properties and light mediators,''
\href{http://doi.org/10.1016/j.physletb.2023.137742}{{Phys. Lett. B} \textbf{839}, 137742 (2023)}.
%[arXiv:2208.06415 [hep-ph]].
%20 citations counted in INSPIRE as of 23 Sep 2023

%\cite{Khan:2023b}
\bibitem{Khan:2023b}
A. N.~Khan,
%``Light new physics and neutrino electromagnetic interactions inXENONnT,''
\href{https://doi.org/10.1016/j.physletb.2022.137650}{Phys. Lett. B \textbf{837}, 137650 (2023)}.
%[arXiv:1908.06045 [hep-ph]].
%40 citations counted in INSPIRE as of 20 Nov 2022



\bibitem{Demirci:2023tui}
M.~Demirci and M.~F.~Mustamin,
%``Solar neutrino constraints on light mediators through coherent elastic neutrino-nucleus scattering,''
\href{https://doi.org/10.1103/PhysRevD.109.015021}{Phys. Rev. D \textbf{109}, 015021 (2024)}.
%[arXiv:2312.17502 [hep-ph]].

\bibitem{DeRomeri:2024dbv}
V.~De Romeri, D.~K.~Papoulias and C.~A.~Ternes,
%``Light vector mediators at direct detection experiments,''
\href{https://doi.org/10.1007/JHEP05(2024)165}{ J. High Energ. Phys. \textbf{2024}, 165 (2024)}.
%[arXiv:2402.05506 [hep-ph]].

\bibitem{Bauer:2018onh}
M.~Bauer, P.~Foldenauer and J.~Jaeckel,
%``Hunting All the Hidden Photons,''
\href{https://doi.org/10.1007/JHEP07(2018)094}{J. High Energ. Phys. \textbf{2018}, 94 (2018)}.
%[arXiv:1803.05466 [hep-ph]].

\bibitem{Lindner:2020kko}
M.~Lindner, Y.~Mambrini, T.~B.~de Melo and F.~S.~Queiroz,
%``XENON1T anomaly: A light Z' from a Two Higgs Doublet Model,''
\href{http://doi.org/10.1016/j.physletb.2020.135972}{Phys. Lett. B \textbf{811}, 135972  (2020)}.
%[arXiv:2006.14590 [hep-ph]].

\bibitem{Bilmis:2015lja}
S.~Bilmis, I.~Turan, T.~M.~Aliev, M.~Deniz, L.~Singh and H.~T.~Wong,
%``Constraints on Dark Photon from Neutrino-Electron Scattering Experiments,''
\href{http://doi.org/10.1103/PhysRevD.92.033009}{Phys. Rev. D \textbf{92}, 033009 (2015)}.
%[arXiv:1502.07763 [hep-ph]].


\bibitem{Fukugita:1986hr}
M.~Fukugita and T.~Yanagida,
%``Baryogenesis Without Grand Unification,''
\href{https://doi.org/10.1016/0370-2693(86)91126-3}{Phys. Lett. B \textbf{174}, 45-47 (1986)}.

\bibitem{Buchmuller:1991ce}
W.~Buchmuller, C.~Greub and P.~Minkowski,
%``Neutrino masses, neutral vector bosons and the scale of B-L breaking,''
\href{https://doi.org/10.1016/0370-2693(91)90952-M}{Phys. Lett. B \textbf{267}, 395-399 (1991)}.

\bibitem{Alves:2015pea}
A.~Alves, A.~Berlin, S.~Profumo and F.~S.~Queiroz,
%``Dark Matter Complementarity and the Z$^\prime$ Portal,''
\href{https://doi.org/10.1103/PhysRevD.92.083004}{Phys. Rev. D \textbf{92}, 083004 (2015)}.
%[arXiv:1501.03490 [hep-ph]].

\bibitem{Allanach:2015gkd}
B.~Allanach, F.~S.~Queiroz, A.~Strumia and S.~Sun,
%``$Z′$ models for the LHCb and $g-2$ muon anomalies,''
\href{https://doi.org/10.1103/PhysRevD.93.055045}{Phys. Rev. D \textbf{93}, 055045 (2016)	[erratum: Phys. Rev. D \textbf{95}, 119902 (2017)]}.
%[arXiv:1511.07447 [hep-ph]].

\bibitem{Allanach:2023uxz}
B.~Allanach and A.~Mullin,
%``Plan B: new Z' models for b \textrightarrow{} s\ensuremath{\ell}$^{+}$\ensuremath{\ell}$^{−}$ anomalies,''
\href{https://doi.org/10.1007/JHEP09(2023)173}{J. High Energ. Phys. \textbf{2023}, 173 (2023)}.
%[arXiv:2306.08669 [hep-ph]].

\bibitem{COHERENT:2017ipa}
D.~Akimov \textit{et al.} (COHERENT Collaboration),
%``Observation of Coherent Elastic Neutrino-Nucleus Scattering,''
\href{https://doi.org/10.1126/science.aao0990}{Science \textbf{357}, 1123-1126 (2017)}.
%[arXiv:1708.01294 [nucl-ex]].

\bibitem{COHERENT:2020iec}
D.~Akimov \textit{et al.} (COHERENT Collaboration),
%``First Measurement of Coherent Elastic Neutrino-Nucleus Scattering on Argon,''
\href{https://doi.org/10.1103/PhysRevLett.126.012002}{Phys. Rev. Lett. \textbf{126}, 012002 (2021)}.
%[arXiv:2003.10630 [nucl-ex]].

\bibitem{COHERENT:2021xmm}
D.~Akimov \textit{et al.} (COHERENT Collaboration),
%``Measurement of the Coherent Elastic Neutrino-Nucleus Scattering Cross Section on CsI by COHERENT,''
\href{https://doi.org/10.1103/PhysRevLett.129.081801}{Phys. Rev. Lett. \textbf{129}, 081801 (2022)}.
%[arXiv:2110.07730 [hep-ex]].

%\cite{Beda:2009kx}
\bibitem{Beda:2009kx}
A.~G.~Beda \textit{et al.} (GEMMA Collaboration), %, E.~V.~Demidova, A.~S.~Starostin, V.~B.~Brudanin, V.~G.~Egorov, D.~V.~Medvedev, M.~V.~Shirchenko and T.~Vylov,
%``GEMMA experiment: Three years of the search for the neutrino magnetic moment,''
\href{http://doi.org/10.1134/S1547477110060063}{{Phys. Part. Nucl. Lett.} \textbf{7}, 406-409 (2010)}.
%[arXiv:0906.1926 [hep-ex]].
%72 citations counted in INSPIRE as of 02 Jan 2023

%\cite{TEXONO:2009knm}
\bibitem{TEXONO:2009knm}
M.~Deniz \textit{et al.} (TEXONO Collaboration),
%``Measurement of Nu(e)-bar -Electron Scattering Cross-Section with a CsI(Tl) Scintillating Crystal Array at the Kuo-Sheng Nuclear Power Reactor,''
\href{http://doi.org/10.1103/PhysRevD.81.072001}{{Phys. Rev. D} \textbf{81}, 072001 (2010)}.
%[arXiv:0911.1597 [hep-ex]].
%224 citations counted in INSPIRE as of 02 Jan 2023

\bibitem{CONNIE:2019xid}
A.~Aguilar-Arevalo \textit{et al.} (CONNIE Collaboration),
%``Search for light mediators in the low-energy data of the CONNIE reactor neutrino experiment,''
\href{http://doi.org/10.1007/JHEP04(2020)054}{J. High Energ. Phys. \textbf{2020}, 54 (2020)}.
%[arXiv:1910.04951 [hep-ex]].

\bibitem{CONUS:2020skt}
H.~Bonet \textit{et al.} (CONUS Collaboration),
%``Constraints on elastic neutrino nucleus scattering in the fully coherent regime from the CONUS experiment,''
\href{http://doi.org/10.1103/PhysRevLett.126.041804}{Phys. Rev. Lett. \textbf{126}, 041804 (2021)}.
%[arXiv:2011.00210 [hep-ex]].

\bibitem{CONUSplus:2025}
 N. Ackermann \textit{et al.} (CONUS+ Collaboration), First observation of reactor antineutrinos by coherent scattering, 
 \href{https://doi.org/10.48550/arXiv.2501.05206}{arXiv:2501.05206 [hep-ex]}.
 
\bibitem{Colaresi:2022obx}
J.~Colaresi, J.~I.~Collar, T.~W.~Hossbach, C.~M.~Lewis and K.~M.~Yocum,
%``Measurement of Coherent Elastic Neutrino-Nucleus Scattering from Reactor Antineutrinos,''
\href{http://doi.org/10.1103/PhysRevLett.129.211802}{Phys. Rev. Lett. \textbf{129}, 211802  (2022)}.
%[arXiv:2202.09672 [hep-ex]].

%\cite{CHARM-II:1994dzw}
\bibitem{CHARM-II:1994dzw}
P.~Vilain \textit{et al.} (CHARM-II Collaboration),
%``Precision measurement of electroweak parameters from the scattering of muon-neutrinos on electrons,''
\href{http://doi.org/10.1016/0370-2693(94)91421-4}{{Phys. Lett. B} \textbf{335}, 246-252  (1994)}.
%283 citations counted in INSPIRE as of 02 Jan 2023

%\cite{LSND:2001akn}
\bibitem{LSND:2001akn}
L.~B.~Auerbach \textit{et al.} (LSND Collaboration),
%``Measurement of electron - neutrino - electron elastic scattering,''
\href{http://doi.org/10.1103/PhysRevD.63.112001}{{Phys. Rev. D} \textbf{63}, 112001  (2001)}.
%[arXiv:hep-ex/0101039 [hep-ex]].
%302 citations counted in INSPIRE as of 02 Jan 2023

%\cite{NA64:2022yly}
\bibitem{NA64:2022yly}
Y.~M.~Andreev \textit{et al.} (NA64 Collaboration),
%``Search for a New B-L Z' Gauge Boson with the NA64 Experiment at CERN,''
\href{http://doi.org/10.1103/PhysRevLett.129.161801}{{Phys. Rev. Lett.} \textbf{129}, 161801  (2022)}.
%[arXiv:2207.09979 [hep-ex]].
%5 citations counted in INSPIRE as of 02 Jan 2023


\bibitem{A1:2011yso}
H.~Merkel \textit{et al.} (A1 Collaboration),
%``Search for Light Gauge Bosons of the Dark Sector at the Mainz Microtron,''
\href{http://doi.org/10.1103/PhysRevLett.106.251802}{Phys. Rev. Lett. \textbf{106}, 251802 (2011)}.
%[arXiv:1101.4091 [nucl-ex]].

\bibitem{ALICE:2012aqc}
B.~Abelev \textit{et al.} (ALICE Collaboration),
%``Centrality Dependence of Charged Particle Production at Large Transverse Momentum in Pb--Pb Collisions at $\sqrt{s_{\rm{NN}}} = 2.76$ TeV,''
\href{http://doi.org/10.1016/j.physletb.2013.01.051}{Phys. Lett. B \textbf{720}, 52-62 (2013)}.
%[arXiv:1208.2711 [hep-ex]].

\bibitem{BaBar:2014zli}
J.~P.~Lees \textit{et al.} (BaBar Collaboration),
%``Search for a Dark Photon in $e^+e^-$ Collisions at BaBar,''
\href{http://doi.org/10.1103/PhysRevLett.113.201801}{Phys. Rev. Lett. \textbf{113}, 201801 (2014)}.
%[arXiv:1406.2980 [hep-ex]].

\bibitem{PHENIX:2014duq}
A.~Adare \textit{et al.} (PHENIX Collaboration),
%``Search for dark photons from neutral meson decays in $p + p$ and $d$ + Au collisions at $\sqrt{s_{NN}} =$ 200 GeV,''
\href{http://doi.org/10.1103/PhysRevC.91.031901}{Phys. Rev. C \textbf{91}, 031901 (2015)}.
%[arXiv:1409.0851 [nucl-ex]].


\bibitem{NA482:2015wmo}
J.~R.~Batley \textit{et al.} (NA48/2 Collaboration),
%``Search for the dark photon in $\pi^0$ decays,''
\href{http://doi.org/10.1016/j.physletb.2015.04.068}{Phys. Lett. B \textbf{746}, 178-185 (2015)}.
%[arXiv:1504.00607 [hep-ex]].

\bibitem{Muong-2:2023cdq}
D.~P.~Aguillard \textit{et al.} (Muon g-2 Collaboration),
%``Measurement of the Positive Muon Anomalous Magnetic Moment to 0.20~ppm,''
\href{http://doi.org/10.1103/PhysRevLett.131.161802}{Phys. Rev. Lett. \textbf{131}, 161802 (2023)}.
%[arXiv:2308.06230 [hep-ex]].

\bibitem{Heurtier:2017}
L. Heurtier and Y. Zhang, %"Supernova constraints on massive (pseudo)scalar coupling to neutrinos",
\href{http://doi.org/10.1088/1475-7516/2017/02/042}{{JCAP} \textbf{02}, 042  (2017)}.

%\cite{Blinov:2019gcj}
\bibitem{Blinov:2019gcj}
N.~Blinov, K.~J.~Kelly, G.~Z.~Krnjaic and S.~D.~McDermott,
%``Constraining the Self-Interacting Neutrino Interpretation of the Hubble Tension,''
\href{http://doi.org/10.1103/PhysRevLett.123.191102}{{Phys. Rev. Lett.} \textbf{123}, 191102  (2019)}.
%[arXiv:1905.02727 [astro-ph.CO]].
%163 citations counted in INSPIRE as of 22 Sep 2023

\bibitem{Huang:2018}
G.-y. Huang, T. Ohlsson, and S. Zhou,
%``Observational constraints on secret neutrino interactions from big bang nucleosynthesis,"
\href{https://link.aps.org/doi/10.1103/PhysRevD.97.075009}{{Phys. Rev. D} \textbf{97}, 075009  (2018)}.


\bibitem{Croon:2020lrf}
D.~Croon, G.~Elor, R.~K.~Leane and S.~D.~McDermott,
%``Supernova Muons: New Constraints on $Z$' Bosons, Axions and ALPs,''
\href{http://doi.org/10.1007/JHEP01(2021)107}{ J. High Energ. Phys. \textbf{2021}, 107 (2021)}.
%[arXiv:2006.13942 [hep-ph]].

\bibitem{Escudero:2019gzq}
M.~Escudero, D.~Hooper, G.~Krnjaic and M.~Pierre,
%``Cosmology with A Very Light L$_{\mu}$ \ensuremath{-} L$_{\tau}$ Gauge Boson,''
\href{http://doi.org/10.1007/JHEP03(2019)071}{J. High Energ. Phys. \textbf{2019}, 71 (2019)}.
%[arXiv:1901.02010 [hep-ph]].

\bibitem{LiXu:2023}
S.-P. Li and X.-J. Xu, %"Production rates of dark photons and Z' in the Sun and stellar cooling bounds",
\href{https://dx.doi.org/10.1088/1475-7516/2023/09/009}{JCAP \textbf{09}, 009 (2023)}.

\bibitem{Erler:2013}
J. Erler and S. Su, %The Weak Neutral Current, 
\href{https://doi.org/10.1016/j.ppnp.2013.03.004}{Prog. Part. Nucl. Phys. 71, 119–149 (2013)}

\bibitem{Tomalak:2019ibg}
O.~Tomalak and R.~J.~Hill,
%``Theory of elastic neutrino-electron scattering,''
\href{https://doi.org/10.1103/PhysRevD.101.033006}{Phys. Rev. D \textbf{101}, 033006 (2020)}.

\bibitem{Bahcall:1995}
J. N. Bahcall, M. Kamionkowski, and A. Sirlin, %Solar neutrinos: Radiative corrections in neutrino-electron scattering experiments,
\href{https://doi.org/10.1103/PhysRevD.51.6146}{Phys. Rev. D \textbf{51}, 6146-6158 (1995)}.

\bibitem{ParticleDataGroup:2024cfk}
S.~Navas \textit{et al.} (Particle Data Group),
%``Review of particle physics,''
\href{https://doi.org/10.1103/PhysRevD.110.030001}{Phys. Rev. D \textbf{110}, 030001 (2024)}.


\bibitem{Formaggio:2012cpf}
J.~A.~Formaggio and G.~P.~Zeller,
%``From eV to EeV: Neutrino Cross Sections Across Energy Scales,''
\href{https://doi.org/10.1103/RevModPhys.84.1307}{Rev. Mod. Phys. \textbf{84}, 1307-1341 (2012)}.
%[arXiv:1305.7513 [hep-ex]].


%\cite{Allanach:2019}
\bibitem{Allanach:2019} B. C. Allanach, J. Davighi and S. Melville, 
%An anomaly-free ATLAS: charting the space of flavour-dependent gauged U(1) extensions of the Standard Model, 
\href{https://doi.org/10.1007/JHEP02(2019)082}{ J. High Energ. Phys. \textbf{2019}, 82 (2019)} [erratum: \href{https://doi.org/10.1007/JHEP08(2019)064}{ J. High Energ. Phys. \textbf{2019}, 64 (2019)}].

\bibitem{Ballet:2019}
P. Ballett, M. Hostert, S. Pascoli,  Y. F. Perez-Gonzalez, Z. Tabrizi, and  R. Z. Funchal,
 %${Z}^{\ensuremath{'}}\mathrm{s}$ in neutrino scattering at DUNE,
\href{https://doi.org/10.1103/PhysRevD.100.055012}{Phys. Rev. D \textbf{100}, 055012 (2019)}.

\bibitem{Barranco:2011wx}
J.~Barranco, A.~Bolanos, E.~A.~Garces, O.~G.~Miranda and T.~I.~Rashba,
%``Tensorial NSI and Unparticle physics in neutrino scattering,''
\href{https://doi.org/10.1142/S0217751X12501473}{{Int. J. Mod. Phys. A} \textbf{27}, 1250147 (2012)}. %[arXiv:1108.1220 [hep-ph]].

\bibitem{Foot:1992ui}
R.~Foot, H.~Lew and R.~R.~Volkas,
%``Electric charge quantization,''
\href{https://doi.org/10.1088/0954-3899/19/3/005}{J. Phys. G: Nucl. Part. Phys \textbf{19}, 361-372 (1993) [erratum: J. Phys. G: Nucl. Part. Phys \textbf{19}, 1067 (1993)]}.
%[arXiv:hep-ph/9209259 [hep-ph]].

\bibitem{Alvarez-Gaume:1983ihn}
L.~Alvarez-Gaume and E.~Witten,
%``Gravitational Anomalies,''
\href{https://doi.org/10.1016/0550-3213(84)90066-X}{Nucl. Phys. B \textbf{234}, 269 (1984)}.


\bibitem{Altmannshofer:2019}
W. Altmannshofer, S. Gori, J. Mart\'{\i}n-Albo, A. Sousa, and M. Wallbank,
%``Neutrino tridents at DUNE,''
\href{https://doi.org/10.1103/PhysRevD.100.115029}{Phys. Rev. D \textbf{100}, 115029 (2019)}.


\bibitem{Chen:2016eab}
J.~W.~Chen, H.~C.~Chi, C.~P.~Liu and C.~P.~Wu,
%``Low-energy electronic recoil in xenon detectors by solar neutrinos,''
\href{https://doi.org/10.1016/j.physletb.2017.10.029}{Phys. Lett. B \textbf{774}, 656-661 (2017)}.
%[arXiv:1610.04177 [hep-ex]].

\bibitem{xraydata:2009}
A.~Thompson \textit{et al.},
%``Review of particle physics,''
\href{https://xdb.lbl.gov/}{X-ray data booklet (2009)}.

\bibitem{Kouzakov:2017hbc}
K.~A.~Kouzakov and A.~I.~Studenikin,
%``Electromagnetic properties of massive neutrinos in low-energy elastic neutrino-electron scattering,''
\href{https://doi.org/10.1103/PhysRevD.95.055013}{Phys. Rev. D \textbf{95}, 055013 (2017) [erratum: Phys. Rev. D \textbf{96}, 099904 (2017)]}.
%[arXiv:1703.00401 [hep-ph]].

\bibitem{Hsieh:2019hug}
C.~C.~Hsieh, L.~Singh, C.~P.~Wu, J.~W.~Chen, H.~C.~Chi, C.~P.~Liu, M.~K.~Pandey and H.~T.~Wong,
%``Discovery potential of multiton xenon detectors in neutrino electromagnetic properties,''
\href{https://doi.org/10.1103/PhysRevD.100.073001}{Phys. Rev. D \textbf{100}, 073001 (2019)}.
%[arXiv:1903.06085 [hep-ph]].

\bibitem{Maltoni:2015kca}
M.~Maltoni and A.~Y.~Smirnov,
%``Solar neutrinos and neutrino physics,''
\href{https://doi.org/10.1140/epja/i2016-16087-0}{Eur. Phys. J. A \textbf{52} no.4, 87 (2016)}.
%[arXiv:1507.05287 [hep-ph]].

\bibitem{Esteban:2020cvm}
I.~Esteban, M.~C. Gonzalez-Garcia, M.~Maltoni, T.~Schwetz, and A.~Zhou, 
%The fate of hints: updated global analysis of three-flavor neutrino oscillations,
\href{http://dx.doi.org/10.1007/JHEP09(2020)178}{{J. High Energ. Phys}  \textbf{2020}, 178 (2020)}.

\bibitem{Cui:2020bwf}
X.~Cui, Z.~Wang, Y.~Ju, X.~Wang, H.~Liu, W.~Ma, J.~Liu, L.~Zhao, X.~Ji and S.~Li, \textit{et al.}
%``Design and commissioning of the PandaX-4T cryogenic distillation system for krypton and radon removal,''
\href{http://dx.doi.org/10.1088/1748-0221/16/07/P07046}{JINST \textbf{16}, P07046 (2021)}.
%[arXiv:2012.02436 [physics.ins-det]].

\bibitem{AtzoriCorona:2022jeb}
M.~Atzori Corona, W.~M.~Bonivento, M.~Cadeddu, N.~Cargioli and F.~Dordei,
%``New constraint on neutrino magnetic moment and neutrino millicharge from LUX-ZEPLIN dark matter search results,''
\href{http://dx.doi.org/10.1103/PhysRevD.107.053001}{Phys. Rev. D \textbf{107} (2023), 053001}.
%[arXiv:2207.05036 [hep-ph]].

\bibitem{Baker:1983tu}
S.~Baker and R.~D.~Cousins,
%``Clarification of the Use of Chi Square and Likelihood Functions in Fits to Histograms,''
\href{http://dx.doi.org/10.1016/0167-5087(84)90016-4}{Nucl. Instrum. Meth. \textbf{221}, 437-442 (1984)}.

%\cite{Fogli:2002pt}
\bibitem{Fogli:2002pt}
G.~L.~Fogli, E.~Lisi, A.~Marrone, D.~Montanino and A.~Palazzo,
%``Getting the most from the statistical analysis of solar neutrino oscillations,''
\href{http://doi.org/10.1103/PhysRevD.66.053010}{{Phys. Rev. D} \textbf{66}, 053010 (2002)}.
%[arXiv:hep-ph/0206162 [hep-ph]].
%405 citations counted in INSPIRE as of 01 Jun 2023

\bibitem{Bahcall:1989ks}
J.~N.~Bahcall, Neutrino Astrophysics, \href{https://www.cambridge.org/tr/universitypress/subjects/physics/astrophysics/neutrino-astrophysics}{Cambridge University Press (1989), ISBN: 9780521379755}.
%``NEUTRINO ASTROPHYSICS,''

\end{thebibliography}
\end{document}